\documentclass[11pt]{iopart}

\usepackage[british]{babel}
\usepackage{color}
\usepackage{mathrsfs}
\usepackage{setstack}
\usepackage{iopams}
\usepackage[T1]{fontenc}
\usepackage{graphicx}

\begin{document}

\title[Random diffusivity from stochastic equations]{Random diffusivity from
stochastic equations: comparison of two models for Brownian yet non-Gaussian
diffusion}

\author{Vittoria Sposini$^{\dagger,\ddagger}$, Aleksei V. Chechkin$^{\dagger,
\flat}$, Flavio Seno$^{\sharp}$, Gianni Pagnini$^{\ddagger,\S}$, and Ralf
Metzler$^{\dagger}$.}
\address{$\dagger$ Institute for Physics \& Astronomy, University of Potsdam,
14476 Potsdam-Golm, Germany}
\address{$\ddagger$ Basque Center for Applied Mathematics, 48009 Bilbao, Spain}
\address{$\flat$ Akhiezer Institute for Theoretical Physics, 61108 Kharkov, Ukraine}
\address{$\sharp$ INFN, Padova Section and Department of Physics and Astronomy
"Galileo Galilei", University of Padova, 35131 Padova, Italy}
\address{$\S$ Ikerbasque - Basque Foundation for Science, Bilbao, Spain}
\ead{rmetzler@uni-potsdam.de}

\begin{abstract}
A considerable number of systems have recently been reported in which Brownian
yet non-Gaussian dynamics was observed. These are processes characterised by a
linear growth in time of the mean squared displacement, yet the probability
density function of the particle displacement is distinctly non-Gaussian, and
often of exponential (Laplace) shape. This apparently ubiquitous behaviour
observed in very different physical systems has been interpreted as resulting
from diffusion in inhomogeneous environments and mathematically represented
through a variable, stochastic diffusion coefficient. Indeed different models
describing a fluctuating diffusivity have been studied. Here we present a new
view of the stochastic basis describing time dependent random diffusivities
within a broad spectrum of distributions. Concretely, our study is based on
the very generic class of the generalised Gamma distribution. Two models for the
particle spreading in such random diffusivity settings are studied. The first
belongs to the class of generalised grey Brownian motion while the second
follows from the idea of diffusing diffusivities. The two processes exhibit
significant characteristics which reproduce experimental results from different
biological and physical systems. We promote these two physical models for the
description of stochastic particle motion in complex environments.
\end{abstract}

\section{Introduction}

The systematic study of the diffusive motion of tracer particles in fluids
dates back to the 19th century, particularly referring to Robert Brown's
experiments observing the erratic motion of granules extracted from pollen grains
which were suspended in water \cite{Brown}. Since then numerous scientists
contributed by improving the experiments \cite{perrin,nordlund,kappler} as well
as in defining the basis of the theory of diffusion \cite{Einstein:Th_diff1,
Sutherland:Th_diff2,Smoluchowski:Th_diff3,Langevin:Th_diff4,vankampen}. In brief,
Brownian or standard diffusion processes are mainly characterised by two central
features: (i) the linear growth in time of the mean-squared displacement (MSD),
\begin{equation}
\label{msd}
\langle x^2(t)\rangle=2Dt,
\end{equation}
where $D$ is the diffusion coefficient, and (ii) the Gaussian probability density
function (PDF) for the particle displacement,
\begin{equation}
\label{gauss}
G(x,t|D)=\frac{1}{\sqrt{4\pi Dt}}\exp\left(-\frac{x^2}{4Dt}\right).
\end{equation}
Here and in the following we focus on a one-dimensional formulation of the model,
a generalisation to higher dimensions can be achieved component-wise.

Discoveries of deviations from the linear time dependence (\ref{msd}) have a
long history. Thus, Richardson already in 1926 reported his famed $t$-cubed
law for the relative particle diffusion in turbulence \cite{richardson}. Scher
and Montroll uncovered anomalous diffusion of the power-law form
\begin{equation}
\label{anomsd}
\langle x^2(t)\rangle\simeq D_{\alpha}t^{\alpha},
\end{equation}
with the anomalous diffusion exponent $0<\alpha<1$ and the generalised diffusion
coefficient $D_{\alpha}$ \cite{Rest_Metz}, for the motion of charge carriers in
amorphous semiconductors \cite{scher}. With the advance of modern microscopy
techniques, in particular, superresolution microscopy, as well as massive
progress in supercomputing, anomalous diffusion of
the type (\ref{anomsd}) has been reported in numerous complex and biological
systems \cite{hoefling,noerregaard}. Thus, subdiffusion with $0<\alpha<1$ was
observed for submicron tracers in the crowded cytoplasm of biological cells
\cite{weiss,elbaum,seisenhuber,Gold-Cox,lene} as well as in artificially
crowded environments \cite{fradin,wong,weiss1,lene1}. Further reports of
subdiffusion come from the motion of proteins embedded in the membranes of
living cells \cite{weiss0,weigel,manzo}. Subdiffusion is also seen in extensive
simulations studies, for instance, of lipid bilayer membranes \cite{kneller,prl,
prx,bba} and relative diffusion in proteins \cite{jeremy}. Superdiffusion, due
to active motion of molecular motors, was observed in various biological cell
types for both introduced and endogenous tracers \cite{elbaum,seisenhuber,nguyen,
selhuber}.

Most of the anomalous diffusion phenomena mentioned
here belong to two main classes of anomalous diffusion: (i) the class of
continuous time random walk processes, in which scale-free power-law waiting
times in between motion events give rise to the law (\ref{anomsd}) \cite{scher,
hughes}, along with a stretched Gaussian displacement probability density $G(x,
t)$ \cite{Rest_Metz,scher,hughes} as well as weak ergodicity breaking and
ageing \cite{Metz_Barkai,johannes}. We note that similar effects of
non-Gaussianity, weak non-ergodicity, and ageing also occur in spatially
heterogeneous diffusion processes \cite{andrey0,andrey1,andrey2,lapeyre}.
(ii) The second one is the class of viscoelastic
diffusion described by the generalised Langevin equation with power-law friction
kernel \cite{lutz,goychuk} and of fractional Brownian motion \cite{fBM}. These
processes are both fuelled by long-range, power-law correlated noise. Its
distribution is Gaussian, so that the displacement probability density $G(x,t)$
is Gaussian, as well. Moreover, these are ergodic processes \cite{lene1,goychuk,
deng,maria,pre12}.

Over the last few years a new class of diffusive processes has been reported,
namely, so-called Brownian yet non-Gaussian diffusion \cite{Wang:BYNG2,Wang:BYNG3}.
This class identifies a dynamics characterised by a linear growth (\ref{msd}) of
the MSD combined with a non-Gaussian probability density function for the particle
displacement. The emergence of a non-Gaussian distribution, despite the Brownian
MSD scaling, suggests the presence of an inhomogeneity that can be located both on
the single tracer particle and on the ensemble levels. The study of these processes
is becoming increasingly relevant with the growing number of complex systems
discovered to exhibit such statistical features. For instance, we mention soft
matter and biological systems, in which the motion of biological macromolecules,
proteins and viruses along lipid tubes and through actin networks \cite{Wang:BYNG2,
Wang:BYNG3}, as well as along membranes and inside colloidal suspension
\cite{Goldstein:BYNG9} and colloidal nanoparticles adsorbed at fluid interfaces
\cite{xue:BYNG8,wang,Dutta-Chak} are studied. We also mention ecological processes,
involving the characterisation of organism movement and dispersal \cite{Hapca:BYNG4,
beta}, as well as processes, that are Brownian but non-Gaussian in certain time
windows of their dynamics. These concern the dynamics of disordered solids, such as
glasses and supercooled liquids \cite{Kob1,Kob:BYNG5,Sciortino} as well as
interfacial dynamics \cite{samanta:BYNG7,skaug}. Also anomalous diffusion processes
of the viscoelastic class that typically are expected to exhibit Gaussian statistic
of displacements, were reported to have non-Gaussian displacements along with
distinct distributions of diffusivity values. These concern the motion of tracer
particles in the cellular cytoplasm \cite{stuhrmann,Lampo:BYNG6,Soares} and the
motion of lipids and proteins in protein-crowded model membranes \cite{prx}.

Here we study two alternative stochastic approaches to non-Gaussian diffusion due
to random diffusivity parameters, namely, generalised grey Brownian motion (ggBM)
and diffusing diffusivities (DD). We analyse their exact behaviour and relate these
approaches to the idea of superstatistics. To prepare the discussion, section
\ref{pathways} presents a primer on the approach of superstatistics and what
has been done in the context of ggBM and DD models. In section \ref{ggBM_sec} we
then study the ggBM model with a random diffusivity distributed according to the
generalised Gamma distribution. In particular, ggBM will be shown to represent a
stochastic description of the superstatistics approach and is equivalent to the
short time limit of the DD model. In section \ref{rand_diff} we formulate a set
of stochastic equations for the dynamics within the DD framework, in which the
diffusivity statistic is governed by the generalised Gamma distribution. This is
then incorporated in the framework of the minimal model of DD in section
\ref{DD_sec}. In section \ref{kurtosis} we describe the behaviour of the kurtosis
of the two models, an important quantity for data analysis. Section \ref{NEIC}
introduces an analysis for an initial non-equilibrium setting for the random
diffusivity, relevant, for instance, for the description of single particle
trajectories. To transfer this concept to the ggBM approach we propose a
non-equilibrium version of ggBM. Finally our conclusions are reported in section
\ref{concl}. In the Appendices some mathematical details are collected.

\section{Pathways to Brownian yet non-Gaussian diffusion: superstatistics
and diffusing diffusivity, and generalised grey Brownian motion}
\label{pathways}

When we talk about an ensemble of particles, we could imagine that non-Gaussian
statistic in this ensemble sense emerges due to the fact that different particles
are located in different environments with different transport characteristics, such
as the diffusion coefficient. If during the observation time each particle remains
in its own environment characterised by a given value $D$ of the diffusivity, the
ensemble of particles shows a mixture of individual Gaussians, weighted by some
distribution $p(D)$ of local diffusivities. This is the idea behind superstatistics,
an approach promoted by Beck and Cohen \cite{Super1}, see also \cite{Superst}. As
a result, the ensemble dynamics is still Brownian yet the PDF of particle
displacements will correspond to a sum or integral of single Gaussians with
specific value of $D$, weighted by the distribution $p(D)$. For instance, an
exponential form for $p(D)$ will produce an exponential shape of the ensemble
displacement PDF, sometimes called a Laplace distribution. We note that there
also exist superstatistical formulations on the basis of the stochastic Langevin 
equation, leading to Brownian yet non-Gaussian behaviour \cite{straeten}. A quite
general superstatistical formulation in terms of the gamma distribution was
put forward by Hapca et al. \cite{Hapca:BYNG4}.

More recently, similar concepts have been sought to describe non-Gaussian
viscoelastic subdiffusion. Thus, Lampo et al.~\cite{Lampo:BYNG6} observed
exponential distributions of the generalised diffusivity $D_{\alpha}$ for
the motion of submicron tracers in living bacteria and eukaryotic cells.
As a theoretical description they used a superstatistical formulation of
the stochastic equation for fractional Brownian motion \cite{Lampo:BYNG6}.
Following the observation of stretched Gaussian shapes of the displacement
PDF in protein-crowded lipid bilayer membranes \cite{prx}, more general
forms for the distribution of the generalised diffusion coefficient were
introduced, see, for instance, \cite{adlampo,Chechkin:DD2}. Viscoelastic,
non-Gaussian diffusion was also described in terms of the generalised
Langevin equation with superstatistical distribution of the friction
amplitude \cite{straeten1,jakub}.

Some other models instead introduce a fluctuating diffusivity, for instance to
describe segregation in solids \cite{Dubinko} or to analyse data from diffusion
processes assessed by modern measurement techniques \cite{Radons}. Brownian
motion in fluctuating environments, or governed by temperature or friction
fluctuations has been studied in \cite{Luczka:1,Luczka:2,Luczka:3} and models
with intermittency between two values of the diffusivity are considered in
\cite{Akimoto1, Akimoto2}. Anomalous diffusion in a disordered system was also
described in terms of a superstatistical model based on a Langevin equation
formulation, combining a Rayleigh-shaped diffusivity distribution with
deterministic power-law growth or decay of the mean diffusivity \cite{andrey}.

A general framework for the description of diffusion in complex environment
is provided also by the class of stochastic processes identified as
generalised grey Brownian motion (ggBM) \cite{lumapa:ggBM, main1:ggBM, main2:ggBM,
Mura:ggBM,daniel}. The basic idea of this approach is that the complexity or
heterogeneity of the medium is completely described by the random nature of
a specific parameter. Choosing this parameter to be the diffusivity leads to
a stochastic interpretation of the system that may be viewed as complementary
to the superstatistics concept and thus suitable for the description of the class
of Brownian yet non-Gaussian processes. We will define ggBM with a random
diffusivity in more detail in the next section \ref{ggBM_sec}, and in the
following demonstrate that ggBM is equivalent to th short time limit of the
DD model.

Recently the idea of DD has received considerable attention. According to this
approach, in addition to the introduction of a population of diffusivities, each
particle during its motion is affected by a continuously changing diffusivity.
Chubynsky and Slater first introduced this model describing the dynamics of the
diffusion coefficient by a biased, stationary random walk with reflecting boundary
conditions \cite{Chubynsky:DD1}. With this assumption the diffusivity changes
slowly step by step, in the short time limit giving rise to normal diffusion
with exponential displacement PDF.\footnote{This approach has some commonalities
in spirit with the correlated continuous time random walk model \cite{vincent,
marcin}.} In the long time regime simulations showed a crossover to Gaussian
diffusion with a single, effective diffusion coefficient \cite{Chubynsky:DD1}.
In a more recent work a direct test of the DD mechanism for diffusion in
inhomogeneous media is reported \cite{Chubynsky_new:DD1}.

The DD concept was further studied by Jain and Sebastian \cite{Sebastian:DD3,
Sebastian:DD3_bis} and Chechkin \etal \cite{Chechkin:DD2}. While Jain and
Sebastian use a path integral approach, Chechkin \etal invoke the concept of
subordination and an explicit exact solution for the PDF in Fourier space.
Despite the different mathematical approach, both models recover the linear
trend of the MSD and a distribution of displacements that at short times is
exponential, while, at long times, it crosses over to a Gaussian with effective
diffusivity, in agreement with the results in \cite{Chubynsky:DD1}. Tyagi and
Cherayil \cite{tyagi:DD5} present a hybrid procedure between the two approaches,
 finding that the modulation of white noise by any stochastic process, whose time
correlation function decays exponentially, is likely to have features similar to
the ones obtained in \cite{Chechkin:DD2,Chubynsky:DD1,Sebastian:DD3,
Sebastian:DD3_bis}. As a recent result we also report the work by Lanoisel{\'e}e and
Grebenkov in which the concept of DD is further investigated, for instance, with
respect to time averages and ergodicity breaking properties \cite{Lanoiselee}.

In this paper we present a detailed comparison between the concept of ggBM with
random diffusivity and the DD model. The main difference between the DD and ggBM
model is represented by the interaction between environment and particles. On
the one hand, in the DD model two different statistical levels are taken into
account, one for the motion of the environment and one for the motion of the
particles. The relation between these two gives rise to specific characteristics.
Thus, at short times the slow variability of the environment guarantees the
superstatistical limit. In the long time regime the diffusivity reaches a
stationary average value leading the particles to develop a Gaussian statistic.
On the other hand, the ggBM model does not directly involve an environment dynamics
but only implies a dynamics in which the statistical features of the environment
continuously drives the particles in their motion, see below for more details.

Concretely, for both ggBM and DD models a set of stochastic equations is introduced
to generate a time dependent random diffusivity with a well defined stationary
distribution. Until now mainly exponential or Gamma distributions have been
considered for the random diffusivity. We here base the discussion on the
generalised Gamma distribution, which represents an even broader class of
distributions including the ones mentioned above, as particular cases. We
define the generalised Gamma distribution by
\begin{equation}
\label{D_PDF}
\gamma^\mathrm{gen}_{\nu,\eta}(D)=\frac{\eta}{D_\star^\nu\,\Gamma(\nu/\eta)}
D^{\nu-1}\exp\left(-\left[\frac{D}{D_\star}\right]^{\eta}\right),
\end{equation}
where $D_\star$ is a positive and dimensional constant and $\nu$ and $\eta$
are positive constants. This distribution encodes the $n$th order stationary
moments
\begin{equation}
\label{moments}
\langle D^n\rangle_{\mathrm{stat}}=D_\star^n\frac{\Gamma([\nu+n]/\eta)}{\Gamma
(\nu/\eta)}.
\end{equation}
The choice of the generalised Gamma distribution is based on experimental
evidence demonstrating its role as a versatile description for generalised
distributions in various complex systems. Indeed, in the context of
superstatistics the generalised Gamma distribution was studied by Beck in
\cite{gengam}. Importantly, the generalised Gamma distribution includes those cases
labelled as Gamma or exponential distribution that have already shown good
agreement with several systems \cite{Hapca:BYNG4,Kob1,Kob:BYNG5,Sciortino}.
Moreover it comprises the cases of stretched
and compressed exponential distributions which may be useful for the
interpretation of various systems \cite{manzo,Hapca:BYNG4,javan,jeon-javan}.

In the following we generalise the ggBM model from references \cite{lumapa:ggBM,
main1:ggBM, main2:ggBM,Mura:ggBM,daniel} to incorporate the generalised Gamma
function (\ref{D_PDF}). We then demonstrate how to reformulate the Ornstein-Uhlenbeck
picture of the DD minimal model \cite{Chechkin:DD2} and the closely related DD
models \cite{Chubynsky:DD1,Sebastian:DD3,Sebastian:DD3_bis} to include the
distribution (\ref{D_PDF}). With this extension both models are considerably more
flexible for the description of measured data. Moreover, we will show that the
ggBM model is a powerful stochastic representation of the superstatistics approach,
and that the ggBM model equals the short time limit of the DD model. Finally, we
consider non-equilibrium conditions in the DD model and propose a non-equilibrium
extension of the ggBM model to consider similar effects in the stochastic setting
of superstatistics. Such non-equilibrium initial conditions represent an important
extension of the random diffusivity models, especially for experimentally relevant
cases of single particle trajectory measurements.

\section{Generalised grey Brownian motion with random diffusivity}
\label{ggBM_sec}

GgBM is defined through the stochastic equation \cite{lumapa:ggBM,main1:ggBM,
main2:ggBM,Mura:ggBM,daniel}
\begin{equation}
\label{ggBM_x}
X_\mathrm{ggBM}(t)=\sqrt{2D}\times W(t),
\end{equation}
for the particle trajectory $X_\mathrm{ggBM}(t)$, in which $W(t)=\int_0^t\xi(t')
dt'$ is standard Brownian motion, the Wiener process defined as the integral over
the white Gaussian noise $\xi(t)$ with zero mean. Moreover, $D$ is a random
diffusivity, here taken to be distributed
according to the generalised Gamma distribution (\ref{D_PDF}). The idea is that
different, but physically identical particles move in disjointed environments, in
which they experience different diffusivities, the essential view of the
superstatistics approach. Alternatively, we could also think of physically
different particles, with different diffusion coefficients, moving in an identical
environment. The latter could, for instance, correspond to an ensemble of tracer
beads with varying radius or different surface properties.

More mathematically speaking, ggBM is defined through the explicit construction of
the underlying probability space based on self-similar increments, and it can be
represented by the stochastic equation $ X_\mathrm{ggBM} =\sqrt{\Lambda}X_g$, where
$\Lambda$ is an independent, non-negative random variable, and $X_g$ is a Gaussian
process \cite{lumapa:ggBM,main1:ggBM,main2:ggBM,Mura:ggBM,daniel}. The
characterisation of this class has also been studied for the case when
$X_g$ is a standard fractional Brownian motion (FBM) and $\Lambda$ is distributed
according to the quite general class of M-Wright functions \cite{Mura:ggBM,
pagnini-paradisi}. We note that the definition (\ref{ggBM_x}) is similar to the
superstatistical Langevin equation models in \cite{straeten,andrey}.

\begin{figure}
\centering
\includegraphics[width=0.48\textwidth]{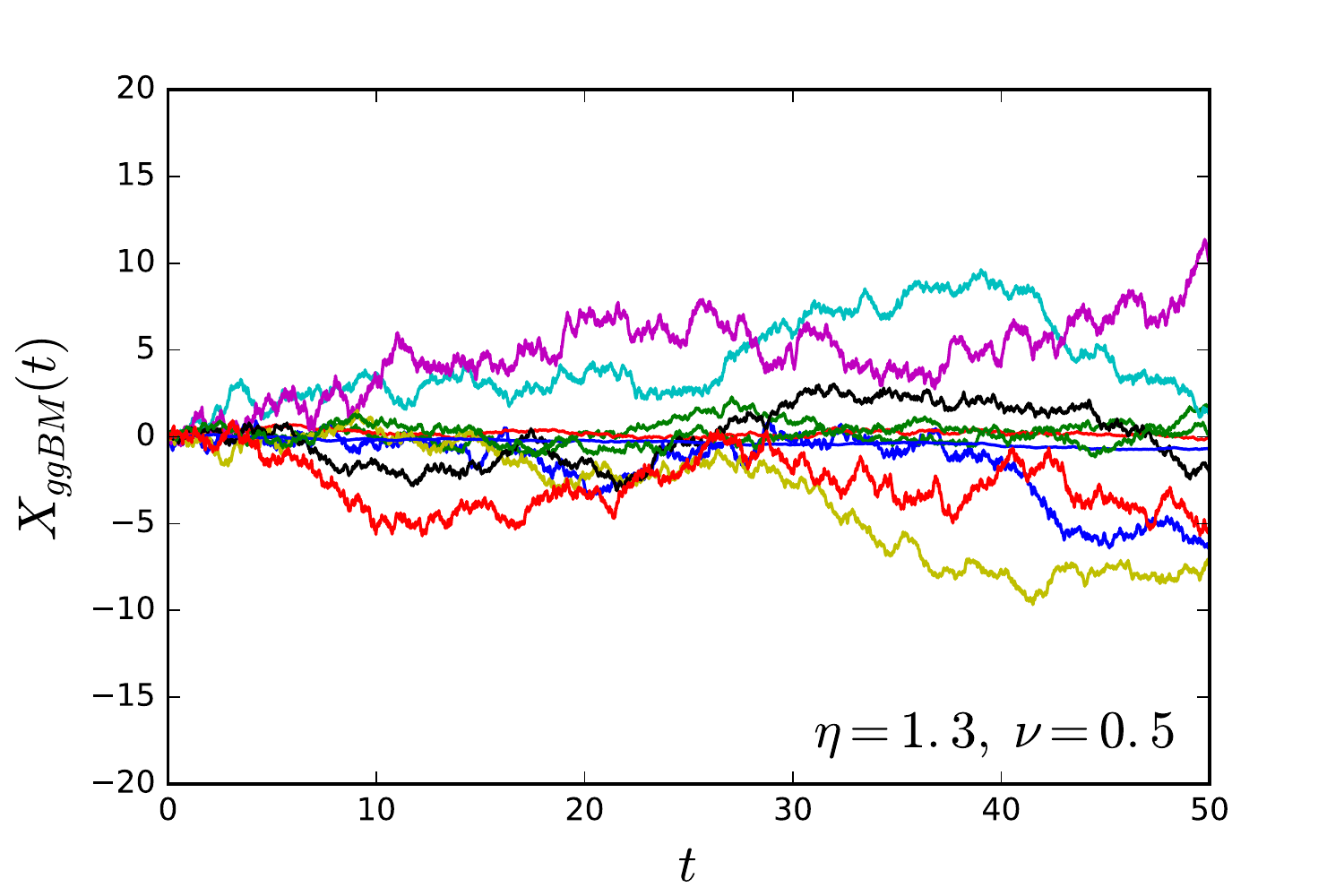}
\includegraphics[width=0.48\textwidth]{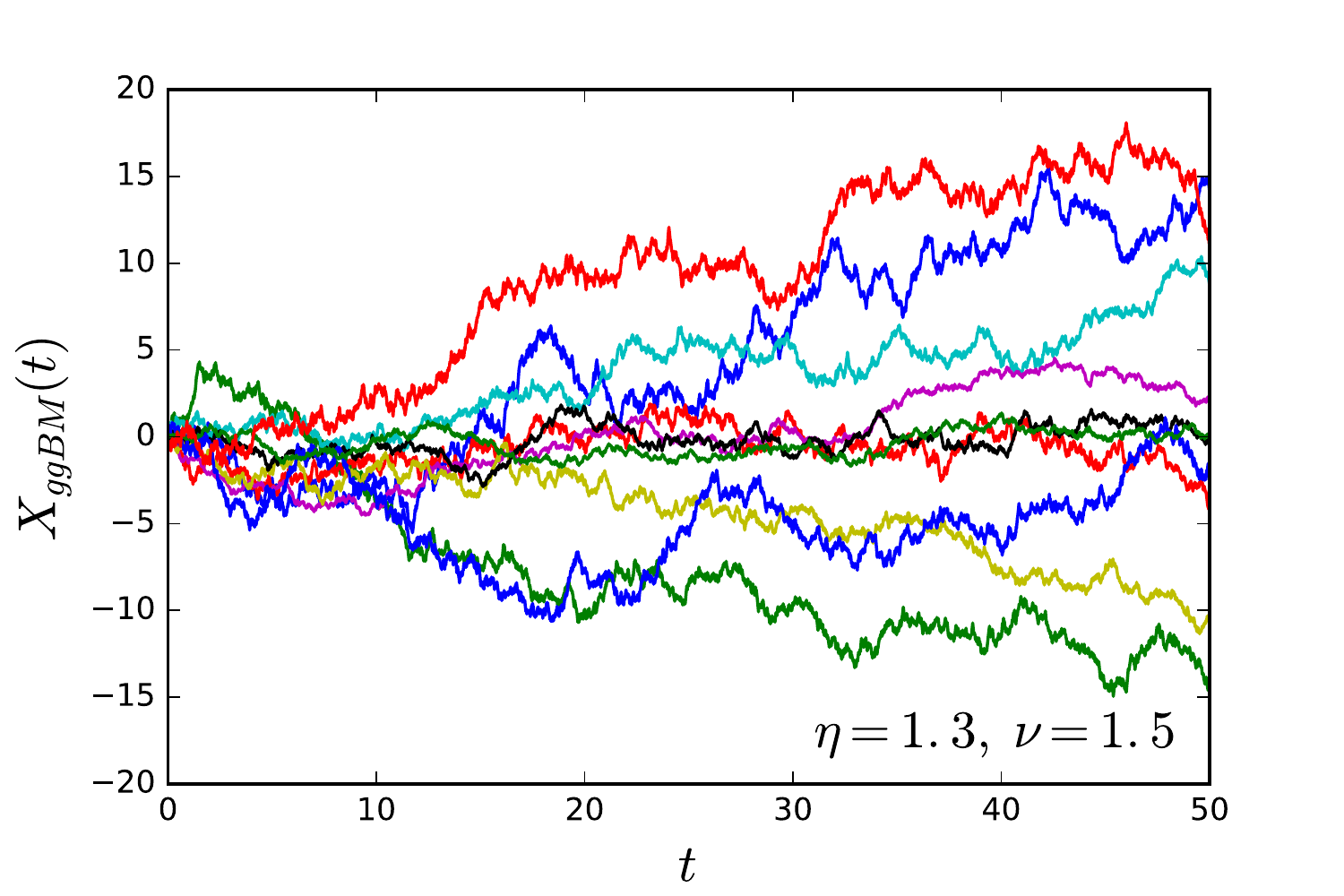}
\includegraphics[width=0.48\textwidth]{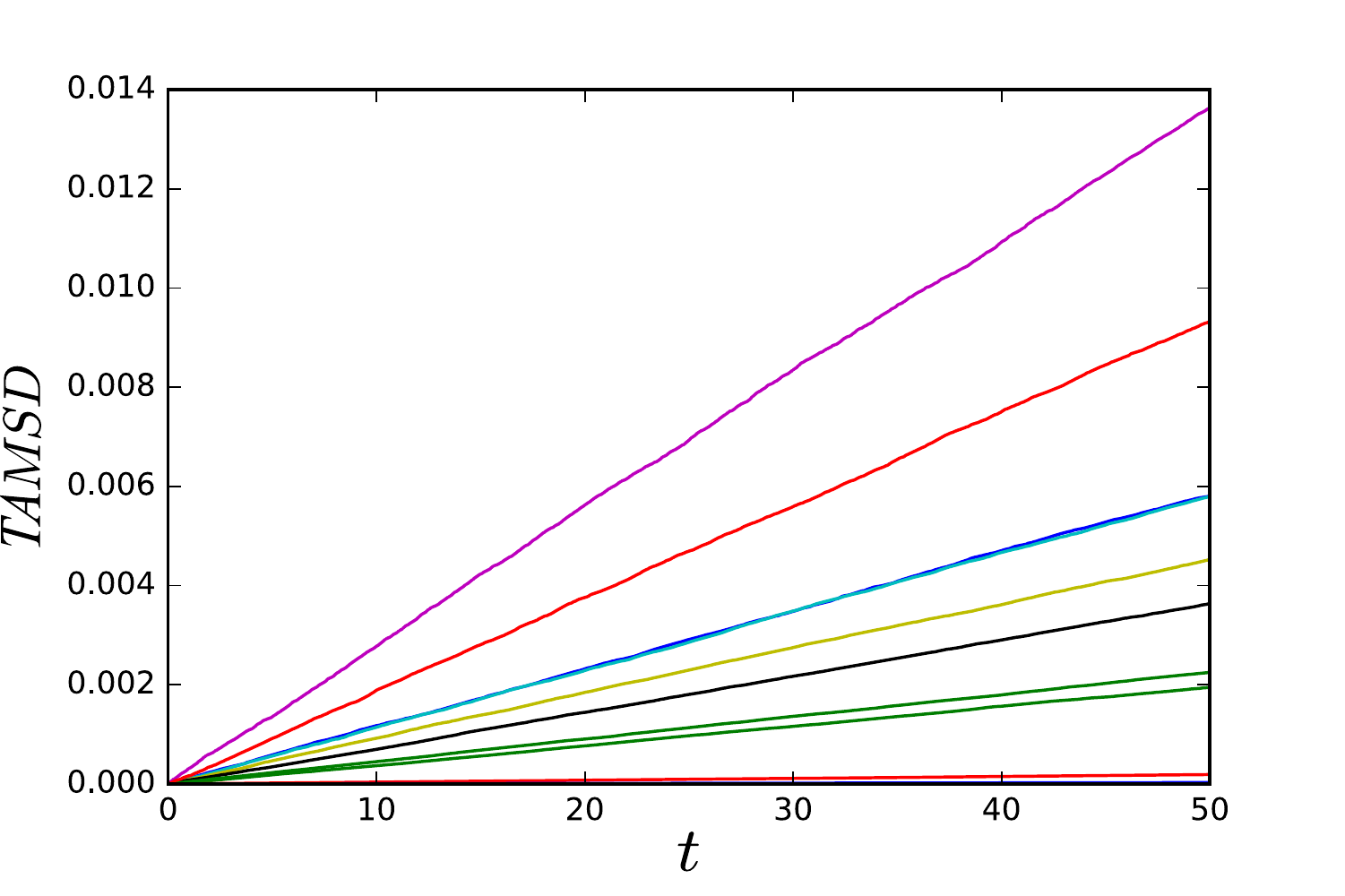}
\includegraphics[width=0.48\textwidth]{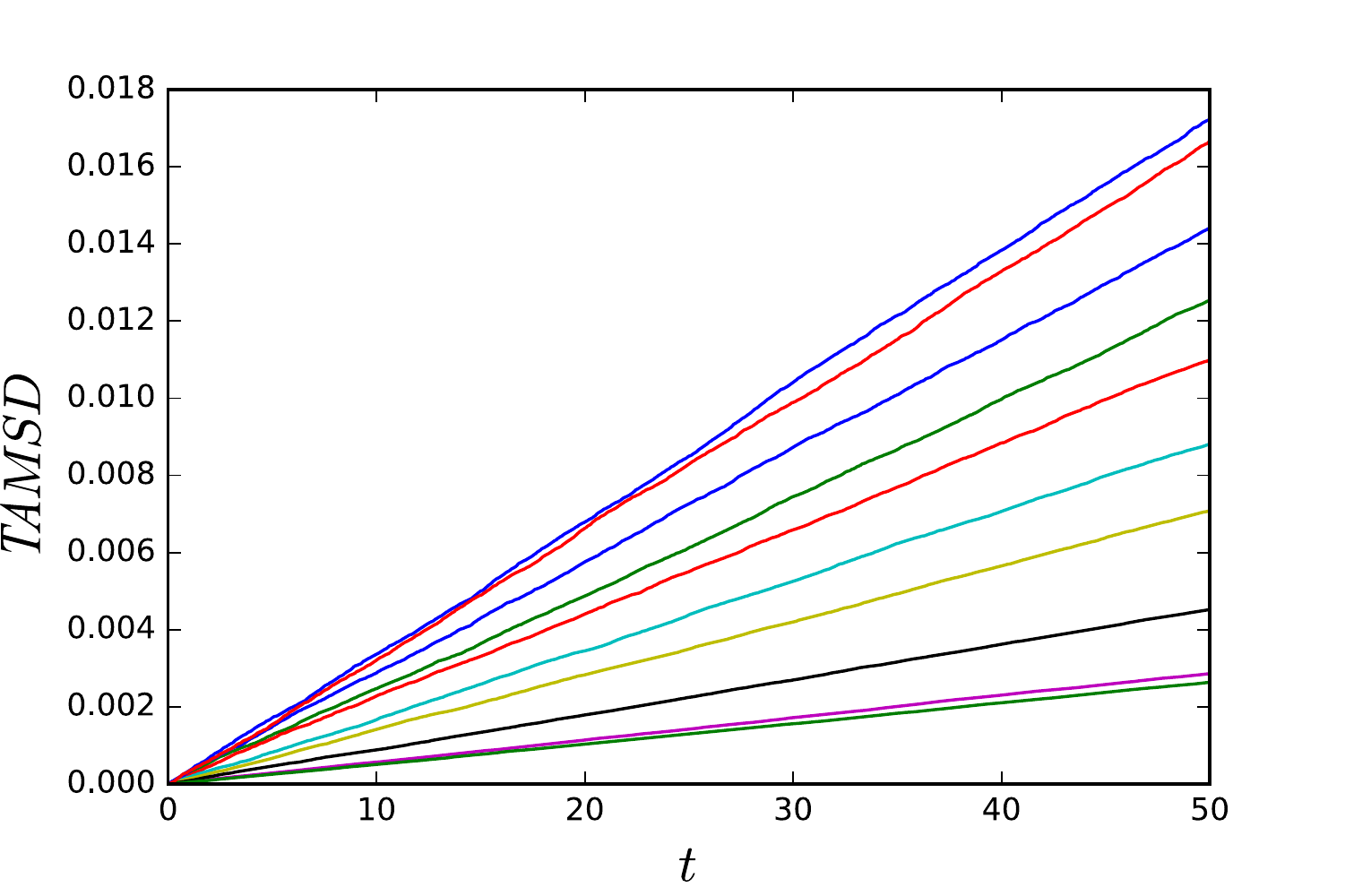}
\caption{Top: Trajectories governed by the ggBM model for $\eta=1.3$ and two
different parameters $\nu$ (see figure legend). Bottom: Time averaged MSD
for the respective traces shown in the top panel, with identical colour coding.
The different trajectories exhibit random diffusivity values and thus random
slopes in the time averaged MSD plots. Within each trajectory the value of $D$
remains fixed.}
\label{traj_ggBM}
\end{figure}

Figure \ref{traj_ggBM} shows trajectories obtained from direct simulations of
the scheme (\ref{ggBM_x}), for which the diffusivity values $D$ are chosen from
the generalised Gamma distribution (\ref{D_PDF}). As a result we obtain a Brownian
motion characterised by a random amplitude, as demonstrated explicitly by the MSD
plots for the same trajectories shown in the bottom panel of figure \ref{traj_ggBM}.
For the value $\nu=1.5$ (right panels) larger $D$ values are observed, in accordance
with the shape of the distribution (\ref{D_PDF}). The ggBM description is indeed
close to the superstatistical concept and fundamentally different from the time
evolution of the sample paths for the DD model, compare figure \ref{traj_DD}.
However, at very short times both processes look much alike, as the DD model at
short times will be shown to reduce to the ggBM model.

The particle displacement distribution can be recovered following Pagnini and
Paradisi \cite{pagnini-paradisi}. If we define with $Z_1$ and $Z_2$ two real
independent random variables whose PDFs are $p_1(z_1)$ and $p_2(z_2)$ with $-
\infty\le z_1\le+\infty$ and $0\le z_2\le+\infty$, respectively, and with the
random variable $Z$ obtained by the product of $Z_1$ and $Z_2^\gamma$, that is,
$Z=Z_1Z_2^\gamma$, then, if we denote the PDF of $Z$ with $p(z)$, we find that
\begin{equation}
p(z)=\int_0^{\infty}p_1\left(\frac{z}{\lambda^\gamma}\right)p_2(\lambda)\frac{d
\lambda}{\lambda^\gamma}.
\label{lemma_2}
\end{equation}

\begin{figure}
\centering
\includegraphics[width=0.49\textwidth]{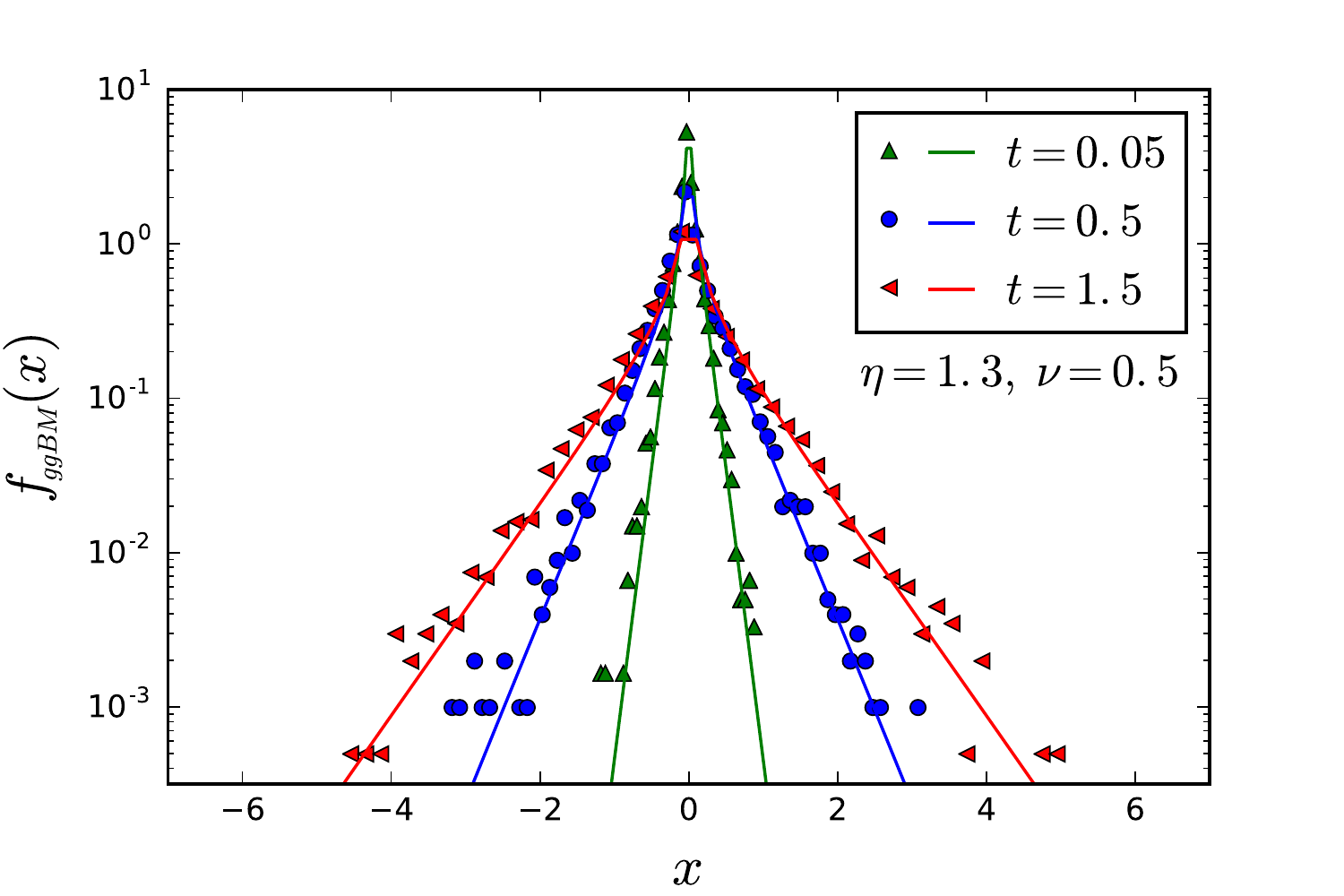}
\includegraphics[width=0.49\textwidth]{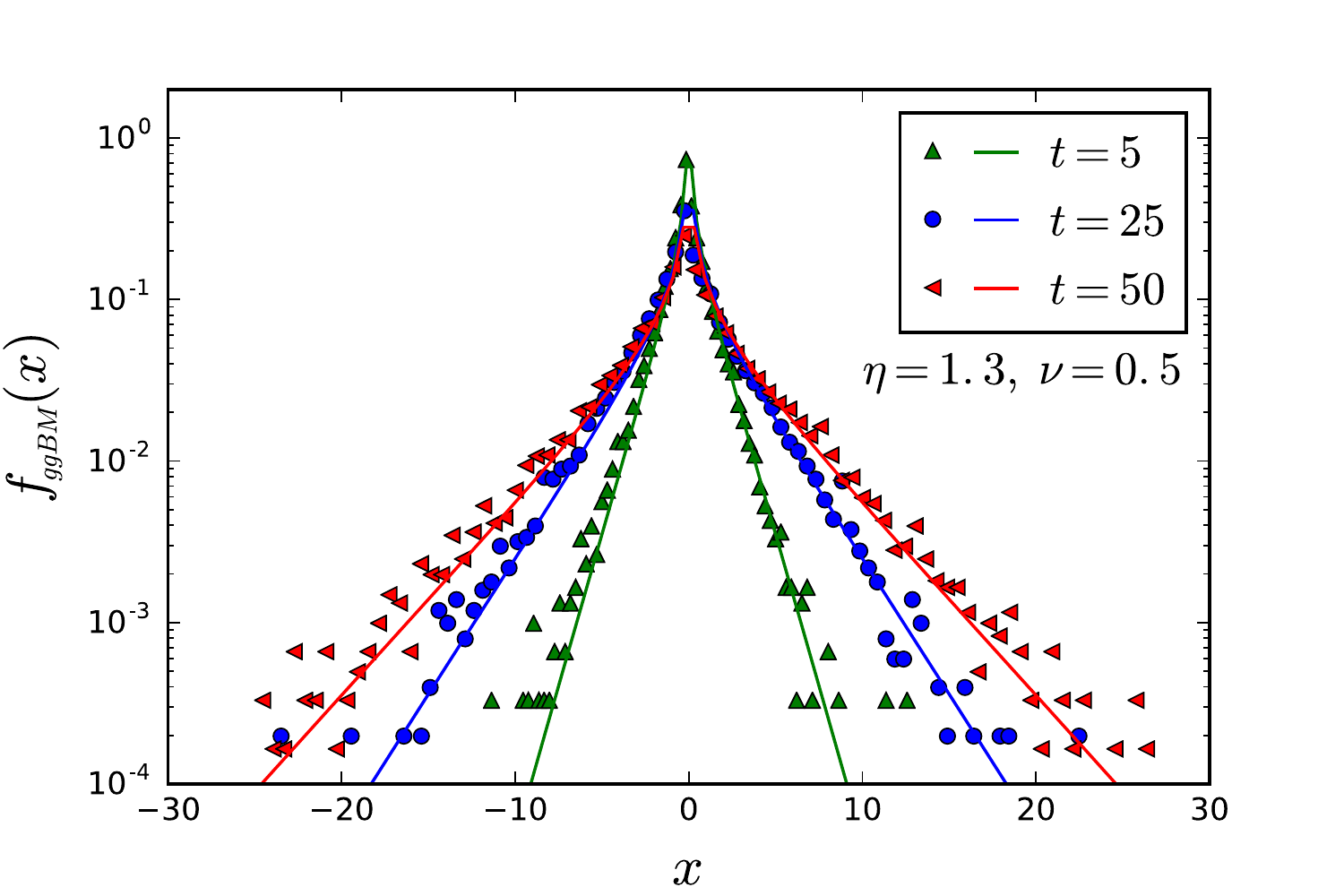}
\caption{Short (a) and long (b) time behaviour of the PDF of the ggBM process for
the parameters $\eta=1.3$ and $\nu=0.5$, as well as $D_{\star}=1/2$. Solid lines
represent the asymptotic behaviour (\ref{asymp_PDF_ggBM_Hfunc}), while symbols
are obtained from stochastic simulations of the ggBM process.}
\label{img_ggBM}
\end{figure}

In the present case we identify $X_{ggBM}(t)$, $W(t)$, and the random diffusivity
$D$ with $Z$, $Z_1$, and $Z_2$, respectively. The PDF for the particle displacement
encoded by equation (\ref{ggBM_x}) and (\ref{lemma_2}) is given by
\begin{eqnarray}
f_\mathrm{ggBM}(x,t)&=&\int_0^{\infty}\frac{1}{\sqrt{2\pi t}}\exp\left(-\frac{
\left(x/\sqrt{2D}\right)^2}{2t}\right)p_D(D)\frac{dD}{\sqrt{2D}}\nonumber\\
&=&\int_0^{\infty}\frac{1}{\sqrt{4\pi Dt}}\exp\left(-\frac{x^2}{4Dt}\right)p_D(D)
dD\nonumber\\
&=&\int_0^{\infty}G(x,t|D)p_D(D)dD,
\label{ggBM_PDF}
\end{eqnarray}
where $G(x,t|D)$ is the Gaussian distribution (\ref{gauss}) for given $D$.
Such a representation of the PDF corresponds to the one of the
superstatistical approach, proving the similarity of the two methods.
The distribution $p_D(D)$ is defined in (\ref{D_PDF}) and the
integral in (\ref{ggBM_PDF}), which can be solved exactly through different methods
(Appendix \ref{app_1}), provides the result (\ref{PDF_ST_Hfunc}) in terms of a Fox
$H$-function (see Appendix \ref{app_1}, where also the series expansion is given).
The asymptotic behaviour of this result acquires the generalised exponential shape
\begin{equation}
\label{asymp_PDF_ggBM_Hfunc}
\fl f_\mathrm{ggBM}(x,t)\sim\frac{(x^2/[4D_\star t])^{(2\nu-\eta-1)/(2[\eta+1])}}{
\Gamma(\nu/\eta)\sqrt{4\pi D_\star t}}\exp\left(-\frac{\eta+1}{\eta}\eta^{\frac{
1}{\eta+1}}\left[\frac{x^2}{4D_\star t}\right]^{\eta/(\eta+1)}\right).
\end{equation}
In particular, the choice $\eta=1$ leads us back to exponential distributions,
with power-law prefactor.
Figure \ref{img_ggBM} demonstrates the agreement between the analytical result
(\ref{asymp_PDF_ggBM_Hfunc}) for the PDF and the result of stochastic simulations of
the underlying ggBM process, for different times and a fixed set of the parameters
$\nu$ and $\eta$. In particular, we see that the shape of the distribution remains
invariant---as for the superstatistical approach---and in contrast to the DD model
analysed below.

\begin{figure}
\centering
\includegraphics[width=0.45\textwidth]{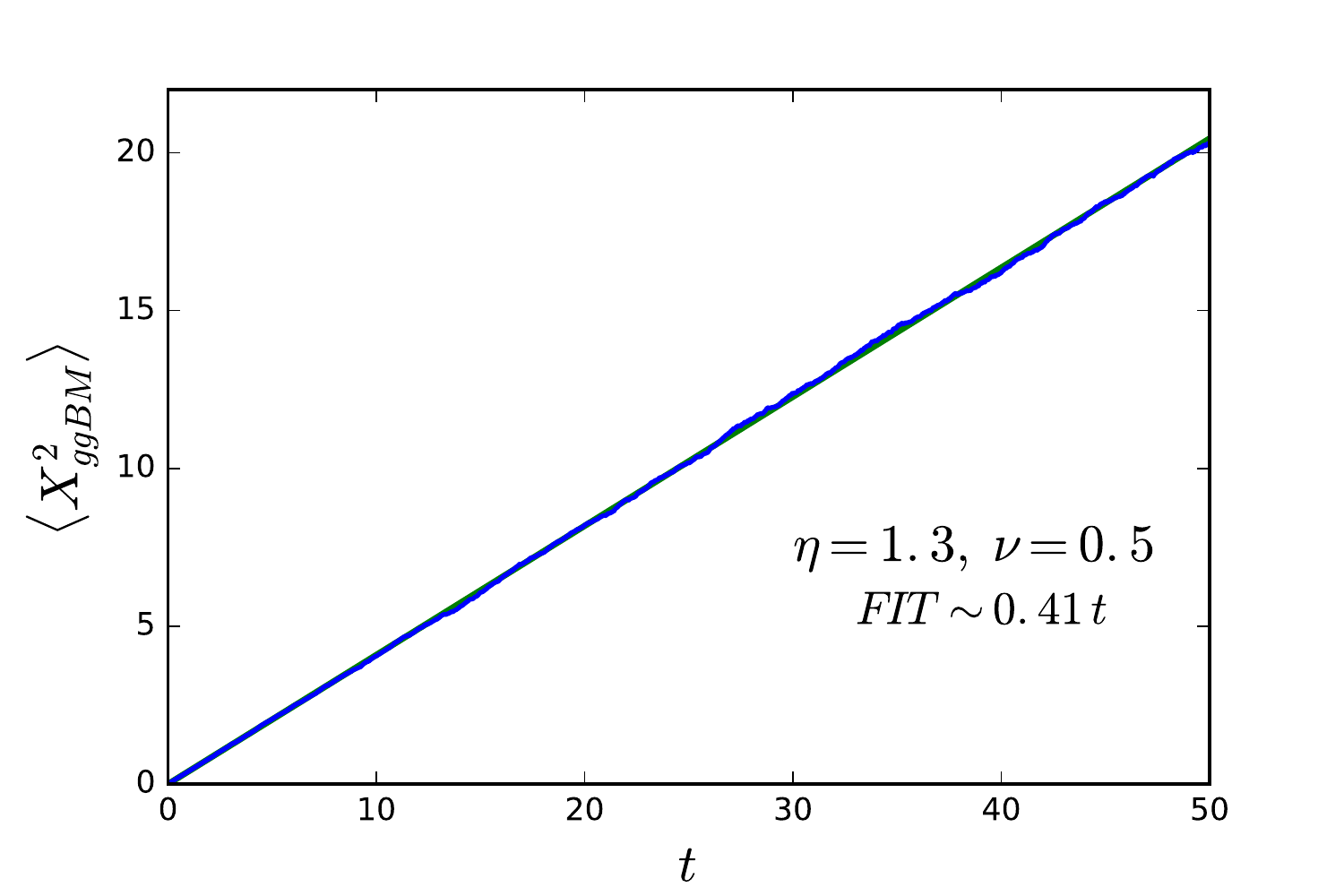}
\caption{Variance of the ggBM model (blue line) and linear fit (solid line).
The corresponding fit parameters are indicated in the figure legend. The
value of $D_{\star}=1/2$.}
\label{2mom_ggBM}
\end{figure}

The MSD follows immediately from the following transformations,
\begin{eqnarray}
\langle X_\mathrm{ggBM}^2(t)\rangle&=&\int_{-\infty}^{\infty} x^2 f_\mathrm{ggBM}
(x,t)dx\nonumber\\
&=&\int_0^{\infty}p_D(D)\int_{-\infty}^{\infty}x^2G(x,t|D)dxdD\nonumber\\
&=&\int_0^{\infty}p_D(D)2DtdD=2t\int_0^{\infty}Dp_D(D)dD\nonumber\\
&=&2\langle D\rangle_\mathrm{stat}t,
\label{2_mom}
\end{eqnarray}
where, according to (\ref{moments}), the effective diffusivity becomes
\begin{equation}
\label{d_eff}
\langle D\rangle_\mathrm{stat}=D_{\star}\Gamma([\nu+1]/\eta)/\Gamma(\nu/\eta).
\end{equation}
Figure \ref{2mom_ggBM} demonstrates the linearity of the variance. The fitted
parameters are consistent with the model prediction, $\langle D\rangle_\mathrm{
stat}=0.20$ comparing to the values chosen in the simulations.

By means of the ggBM approach and with the introduction of a generalised Gamma
distribution for the diffusivity we are able to reproduce a diffusive motion
with a linear scaling of the MSD and a PDF characterised by a stretched or
compressed Gaussian with a power-law prefactor. This is our first main result.

\section{Diffusing diffusivity: stochastic equations for random diffusivity}
\label{rand_diff}

We now consider the diffusion coefficient $D(t)$ to be a random function of
time, defined by means of the auxiliary variable $Y(t)$ through $D(t)=Y^2(t)$,
similarly to the DD minimal model introduced earlier \cite{Chechkin:DD2}.
Our goal is to construct a stochastic equation for the additional variable
$Y(t)$ such that the stationary PDF for its square is the generalised Gamma
distribution in (\ref{D_PDF}). Thus, our present model is represented by the
following set of stochastic equations
\numparts
\begin{eqnarray}
\label{diff_syst}
dY&=&a(Y)dt+\sigma\times dW(t)\\
D(t)&=&Y^2(t),
\end{eqnarray}
\endnumparts
where $a(Y)$ is a nonlinear function whose explicit shape is obtained below,
$\sigma$ is a constant and $W(t)$ is a Wiener process with variance $\langle
W^{^2}(t)\rangle=t$. The physical dimension of the auxiliary variable is
$[Y]=\mathrm{cm}/\mathrm{sec}^{1/2}$ and for the constant $\sigma$ we have
$[\sigma]=\mathrm{cm}/\mathrm{sec}$.

Our approach is based on the central idea that it is possible to establish
a direct relation between the PDFs of the two variables $Y(t)$ and
$D(t)$. This allows us to introduce a completely new dynamics for the
auxiliary variable. Such a dynamics, even though more complex, allows to
reproduce a more general class of PDFs for the random diffusivity and thus
provides a significant extension of the DD model, which will be our second
main result.

To proceed we set $p(Y,t)$ to represent the PDF of the process $Y(t)$ described in
(\ref{diff_syst}). It fulfils the Fokker-Plank equation \cite{vankampen}
\begin{equation}
\frac{\partial p(Y,t)}{\partial t}+\frac{\partial a(Y)p(Y,t)}{\partial Y}=
\frac{\sigma^2}{2}\frac{\partial^2 p(Y,t)}{\partial Y^2}.
\label{F-P}
\end{equation}
Considering the stationary situation the corresponding time independent PDF
$p_Y(Y)$ fulfils the equation
\begin{equation}
\frac{\partial a(Y)p_Y(Y)}{\partial Y}=\frac{\sigma^2}{2}\frac{\partial^2 p_Y(Y)}{
\partial Y^2},
\end{equation}
from which we infer the relation
\begin{equation}
a(Y)=\frac{\sigma^2}{2p_Y(Y)}\frac{\partial p_Y(Y)}{\partial Y},
\label{a_def}
\end{equation}
directly relating the drift coefficient $a(Y)$ with the stationary distribution
of $Y(t)$ \cite{thomson}.

We then recall that, given two random variables $Z_1$ and $Z_2$ related by $Z_2
=g(Z_1)$, for appropriate functions $g(z)$ we have \cite{Ross}
\begin{equation}
p_{Z_2}(z_2)=p_{Z_1}\big( g^{-1}(z_2)\big)\left|\frac{d}{dz_2}g^{-1}(z_2)\right|.
\label{PDF_rel}
\end{equation}
This implies that the distributions of the variables $Y(t)$ and $D(t)$ are related
via
\begin{equation}
p_Y(Y,t)=|Y|p_D(Y^2,t).
\label{PDF_y_D_rel}
\end{equation}

Based on this we construct a set of stochastic equations for the desired quantity
$D(t)$. Starting from the chosen stationary distribution $p_D(D)$ of the random
diffusivity we define the stationary distribution $p_Y(Y)$ for the auxiliary
variable $Y(t)$ by means of equation (\ref{PDF_y_D_rel}). Finally relation
(\ref{a_def}) allows us to recover the suitable coefficient $a(Y)$ in equation
(\ref{diff_syst}). Following the described scheme for the generalised Gamma
distribution (\ref{D_PDF}) we obtain
\begin{equation}
p_Y(Y)=|Y|\frac{\eta}{D_\star^{\nu}\Gamma(\nu/\eta)}Y^{2(\nu-1)}\exp\left(-\left[
\frac{Y}{\sqrt{D_\star}}\right]^{2\eta}\right),
\label{PDF_y}
\end{equation}
and thus
\begin{equation}
\fl\frac{\partial p_Y (Y)}{\partial Y}(Y)=\frac{\eta\mathrm{sgn}(Y)}{D_\star^{\nu}
\Gamma(\nu/\eta)}Y^{2(\nu-1)}\exp\left(-\left[\frac{Y}{\sqrt{D_\star}}\right]^{
2\eta}\right)\left(2\nu-1-2\eta\left[\frac{Y}{\sqrt{D_\star}}\right]^{2\eta}\right).
\end{equation}
This finally leads us to the desired drift coefficient
\begin{equation}
a(Y)=\frac{\sigma^2}{2Y}\left(2\nu-1-2\eta\left[\frac{Y}{\sqrt{D_\star}}\right]^{
2\eta}\right).
\label{a}
\end{equation}
The stochastic equations (\ref{diff_syst}) together with the explicit form
(\ref{a}) of the drift coefficient for the diffusivity fluctuations provide
a complete and generalised analogue of the DD model, which is extremely
flexible for the modelling of experimental data.

We notice that in the particular case of $\nu=0.5$ and $\eta=1$ we recover the
Ornstein-Uhlenbeck model (diffusion in an harmonic potential) considered in the
original minimal DD model \cite{Chechkin:DD2}. As already remarked in
\cite{Chechkin:DD2} in this setting the resulting stochastic equation for $D(t)$ is
nothing else than the Heston model, that is widely used in financial mathematics
and specifies the time evolution of the stochastic volatility of a given asset
\cite{Lanoiselee,heston,yako}.

\begin{figure}
\centering
\includegraphics[width=0.45\textwidth]{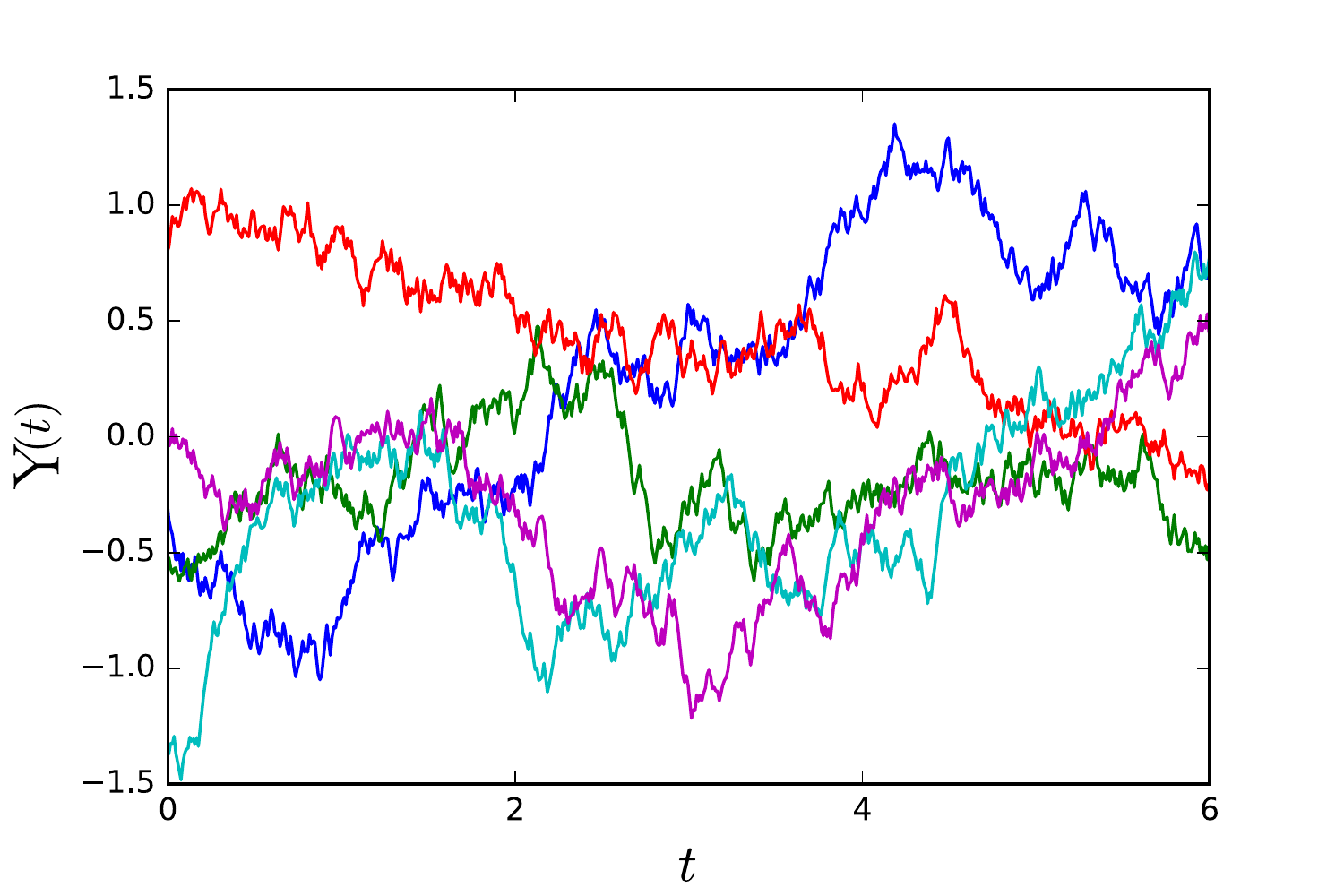}
\includegraphics[width=0.45\textwidth]{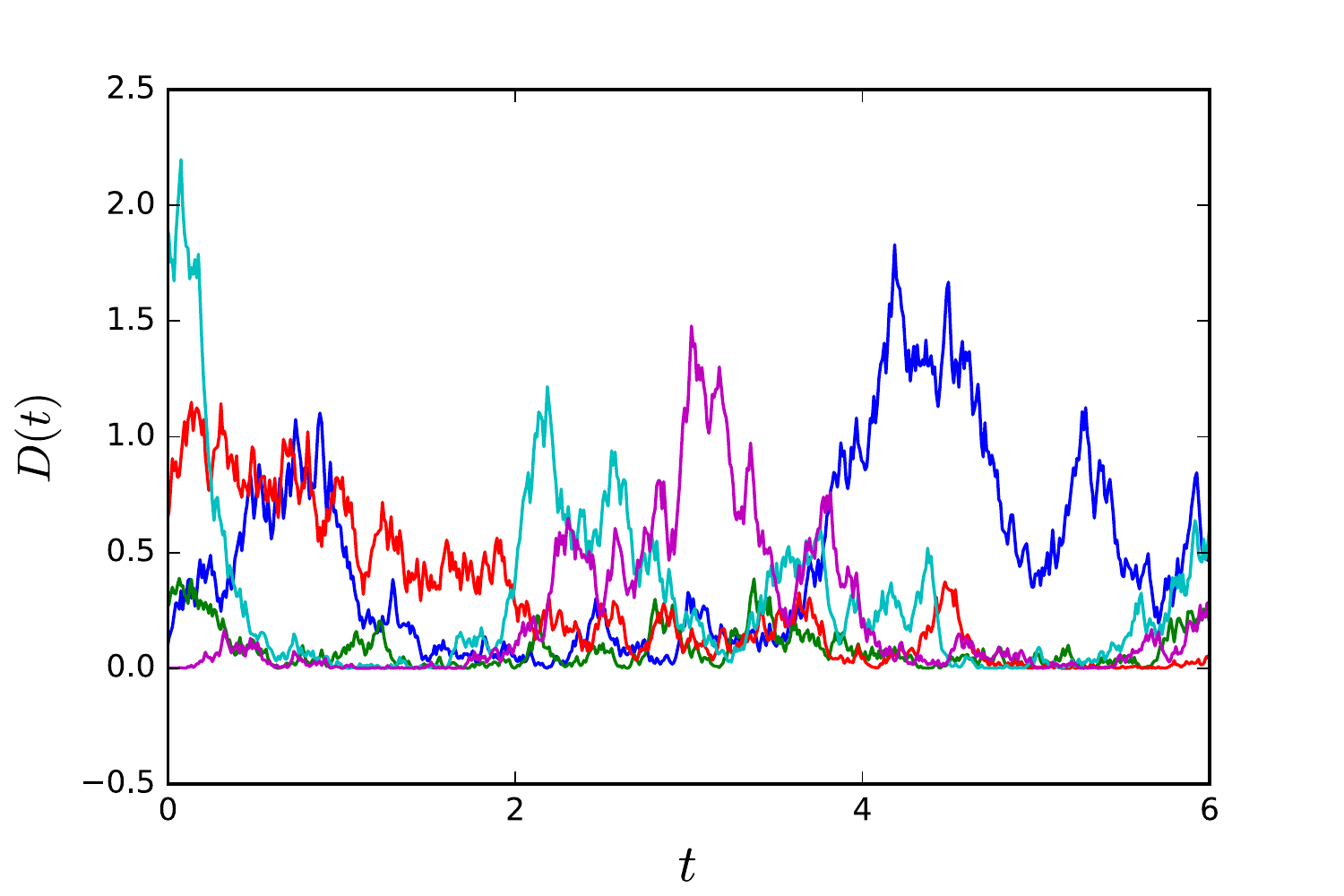}\\
\includegraphics[width=0.45\textwidth]{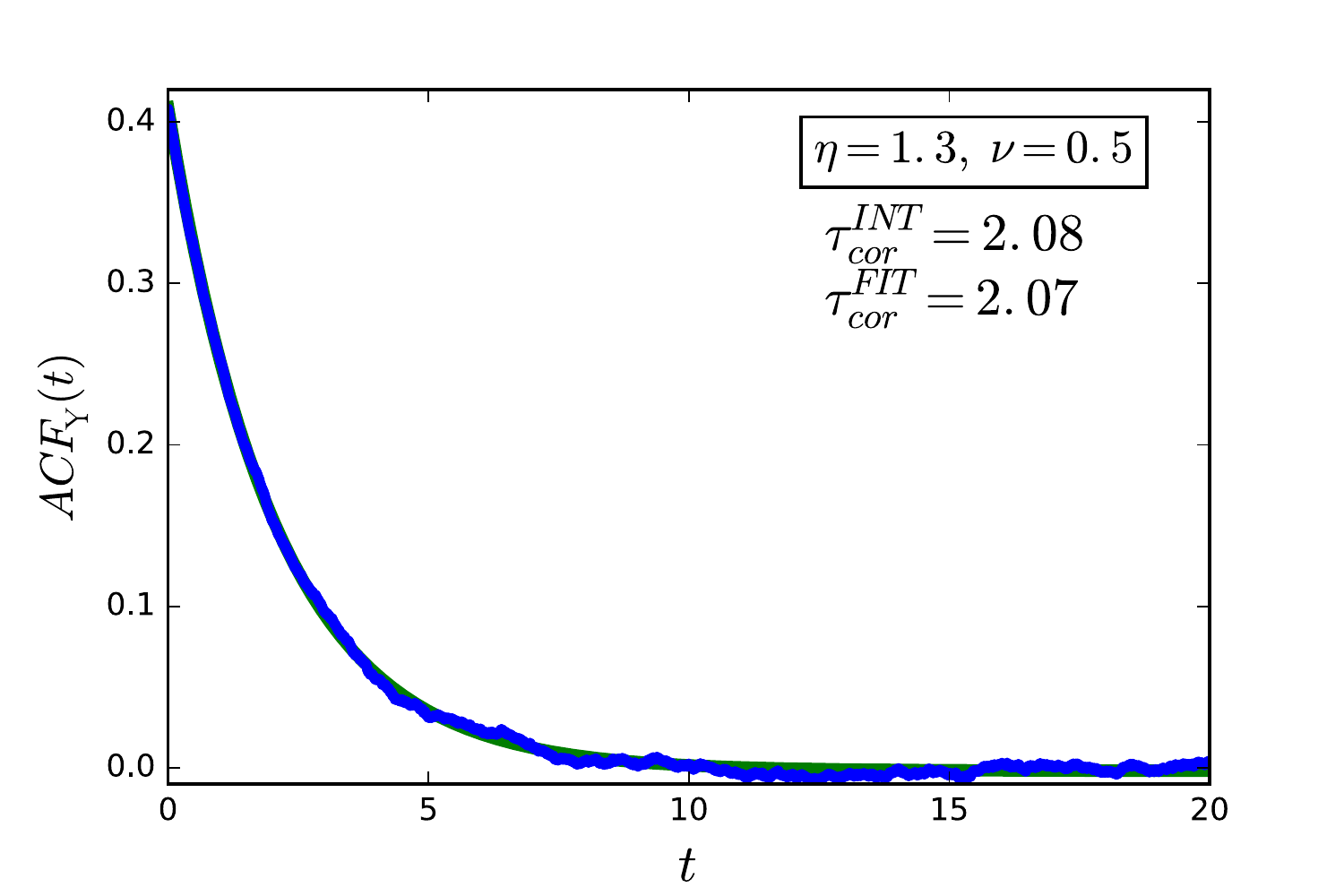}
\includegraphics[width=0.45\textwidth]{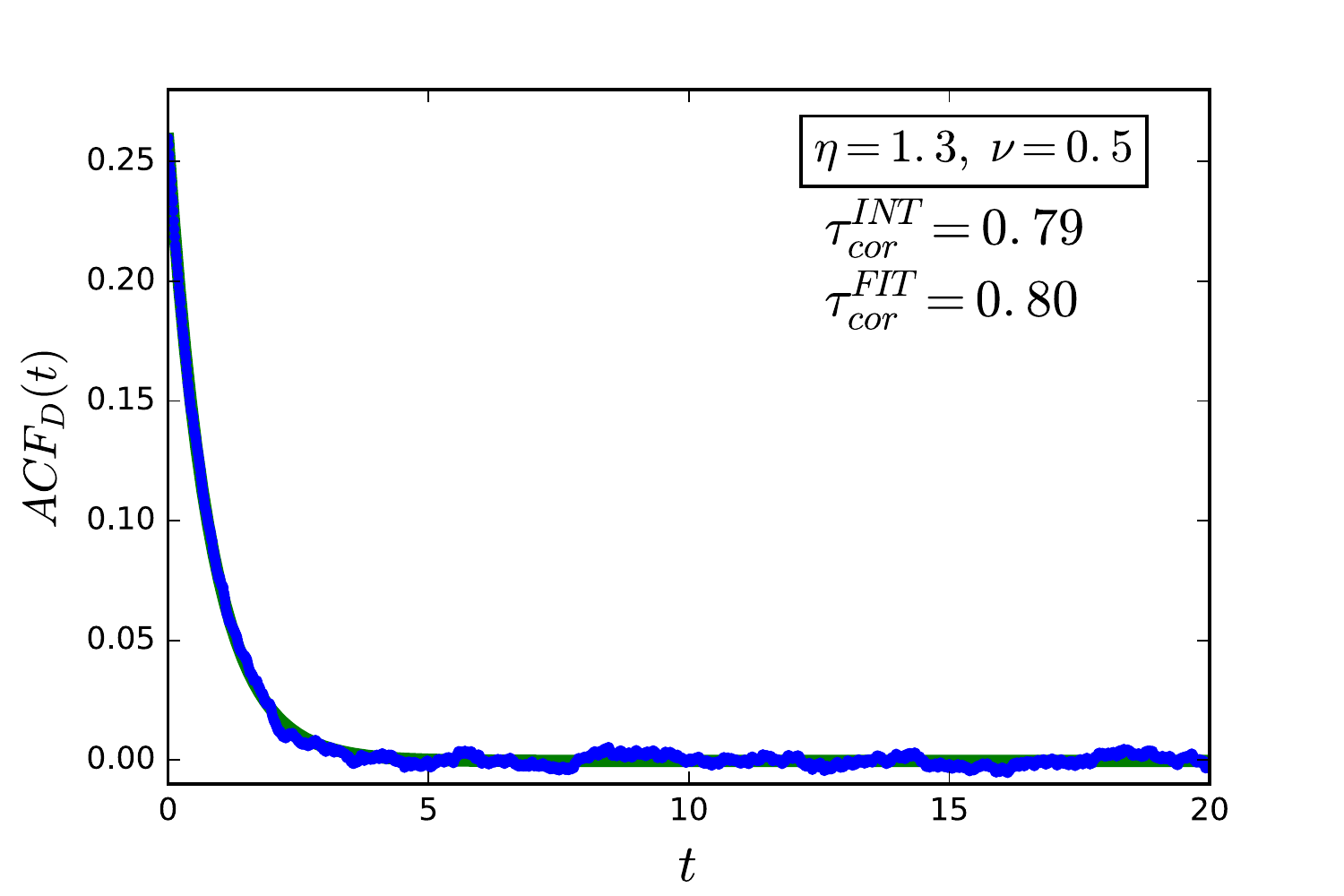}
\caption{Top: Trajectories and bottom: autocorrelation functions (\ref{auto}),
of the auxiliary variable $Y(t)$ and the random diffusivity $D(t)$ in the DD
model. The green solid lines in the autocorrelation function plots represent
exponential fits. We took $\nu=0.5$ and $\eta=1.3$.}
\label{img_rand_diff_1}
\end{figure}

\begin{figure}
\centering
\includegraphics[width=0.45\textwidth]{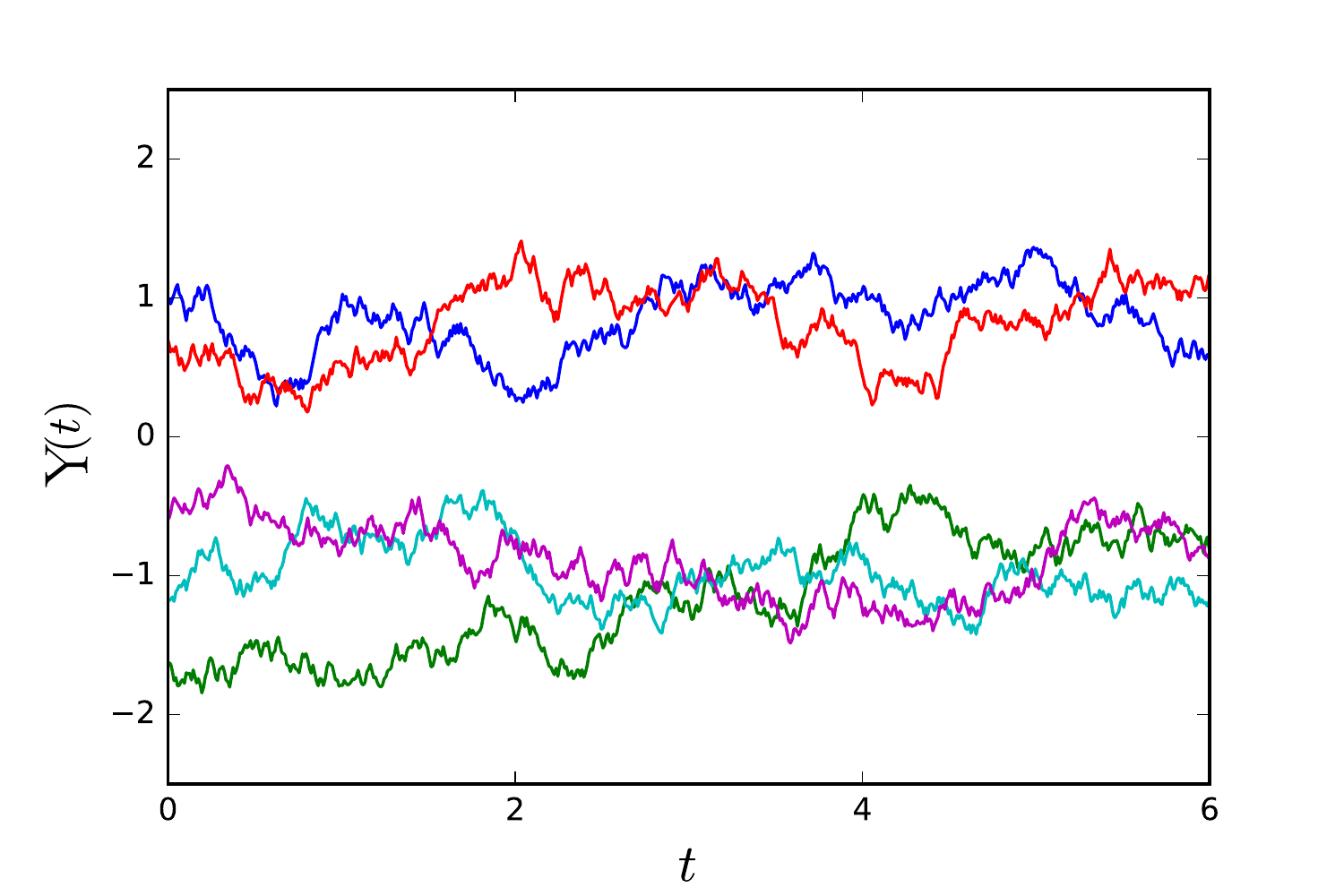}
\includegraphics[width=0.45\textwidth]{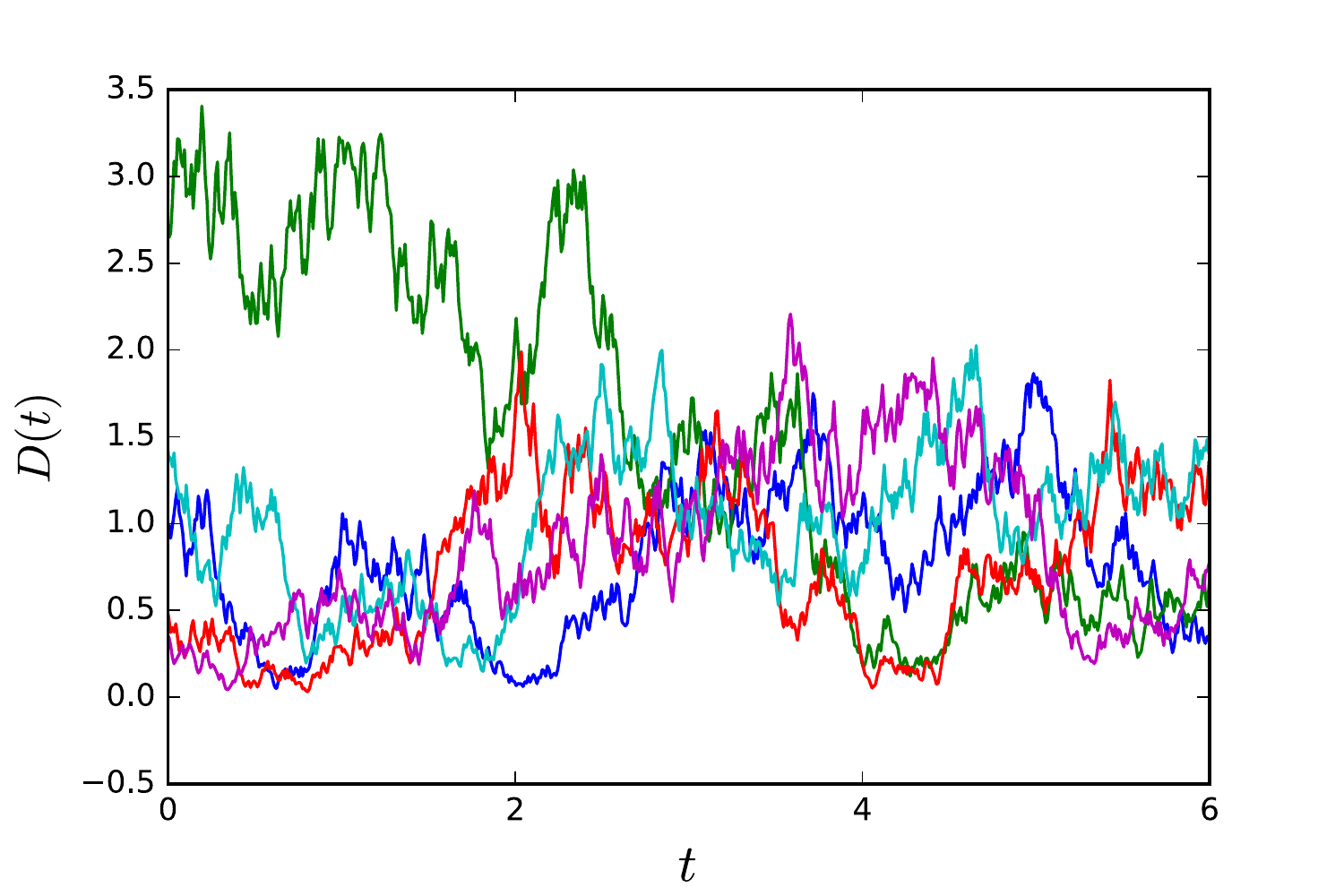}\\
\includegraphics[width=0.45\textwidth]{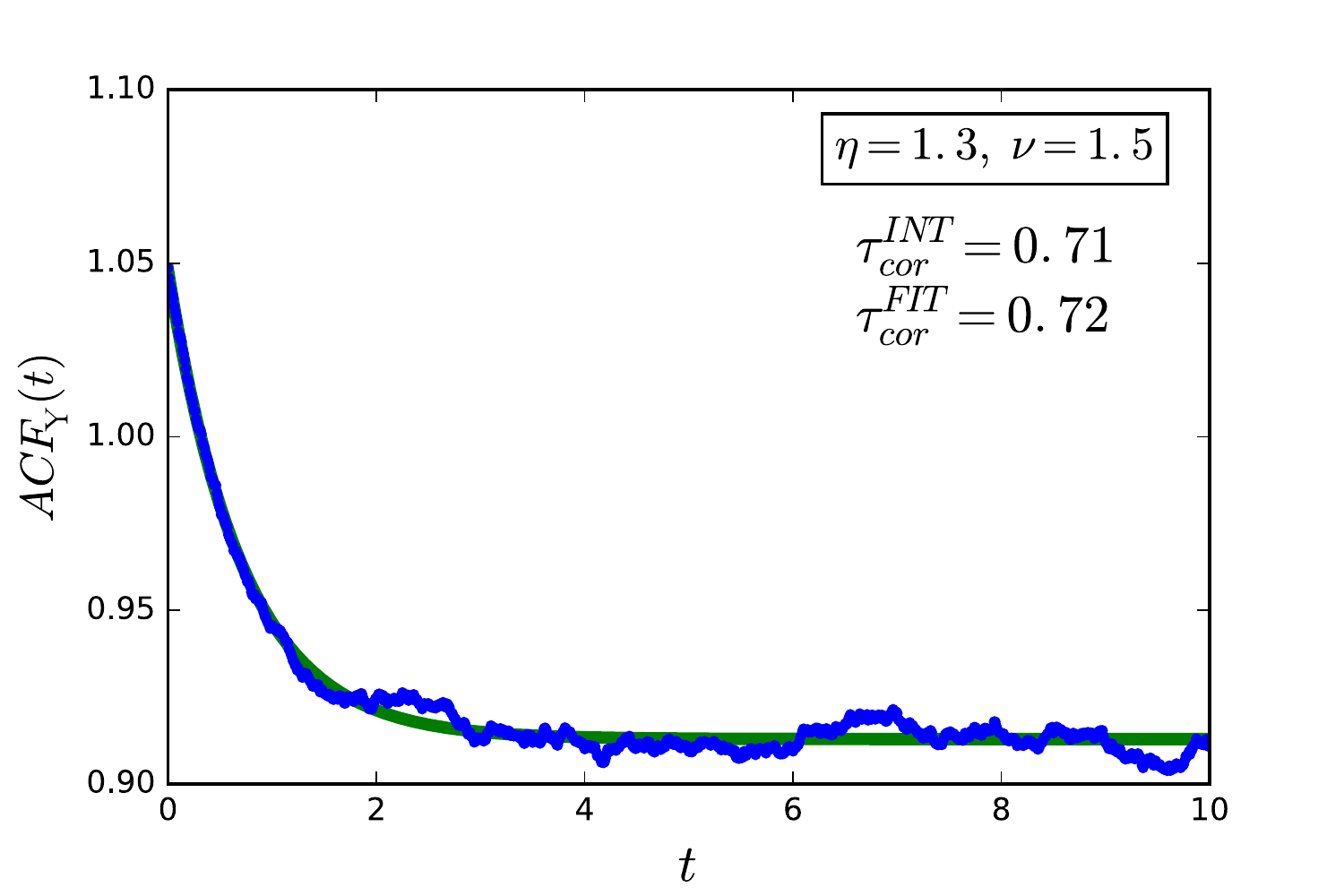}
\includegraphics[width=0.45\textwidth]{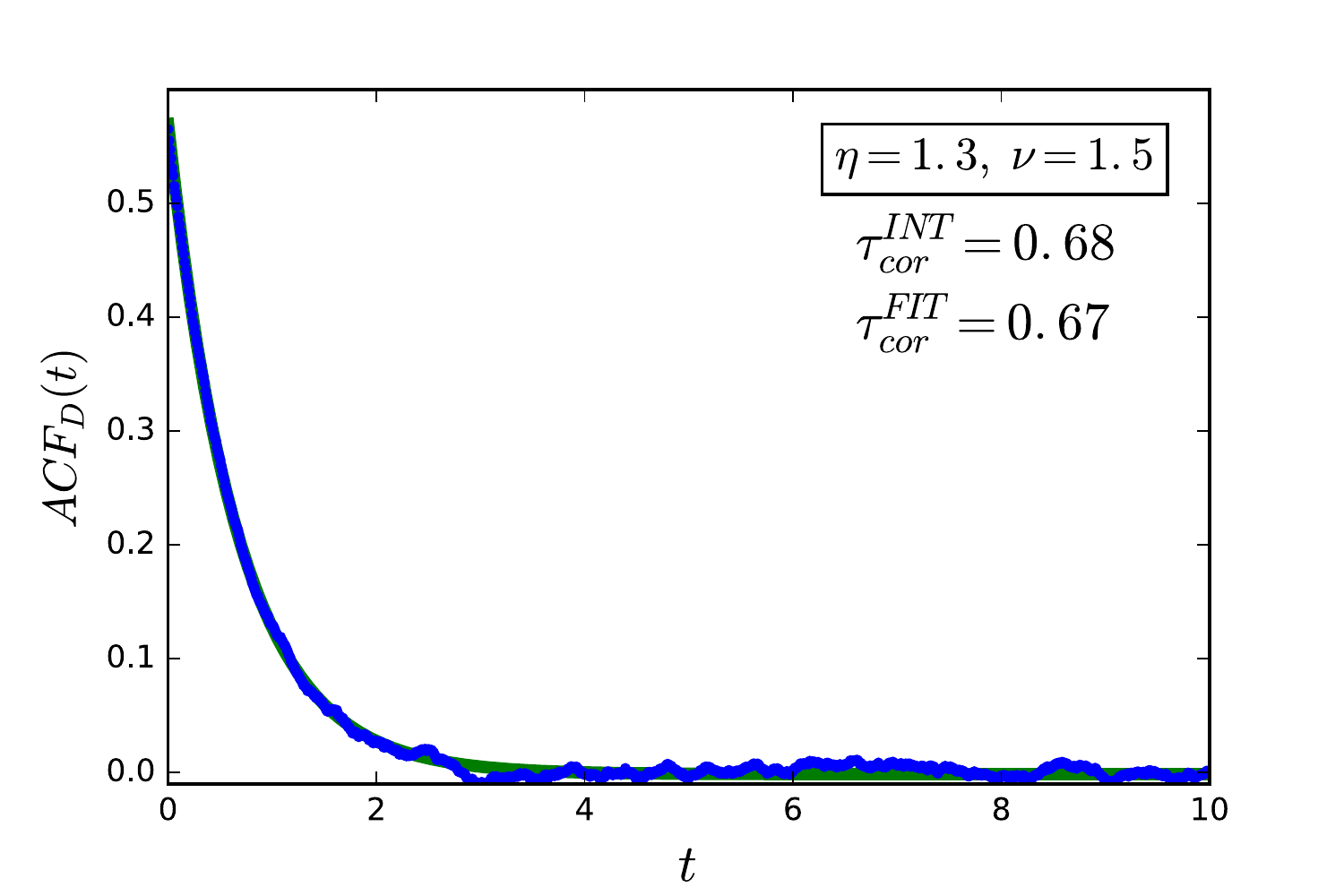}
\caption{Top: Trajectories and bottom: autocorrelation functions (\ref{auto}),
of the auxiliary variable $Y(t)$ and the random diffusivity $D(t)$ in the DD
model. The green solid lines in the autocorrelation function plots represent
exponential fits. We took $\nu=1.5$ and $\eta=1.3$.}
\label{img_rand_diff_2}
\end{figure}

\begin{figure}
\centering
\includegraphics[width=0.49\textwidth]{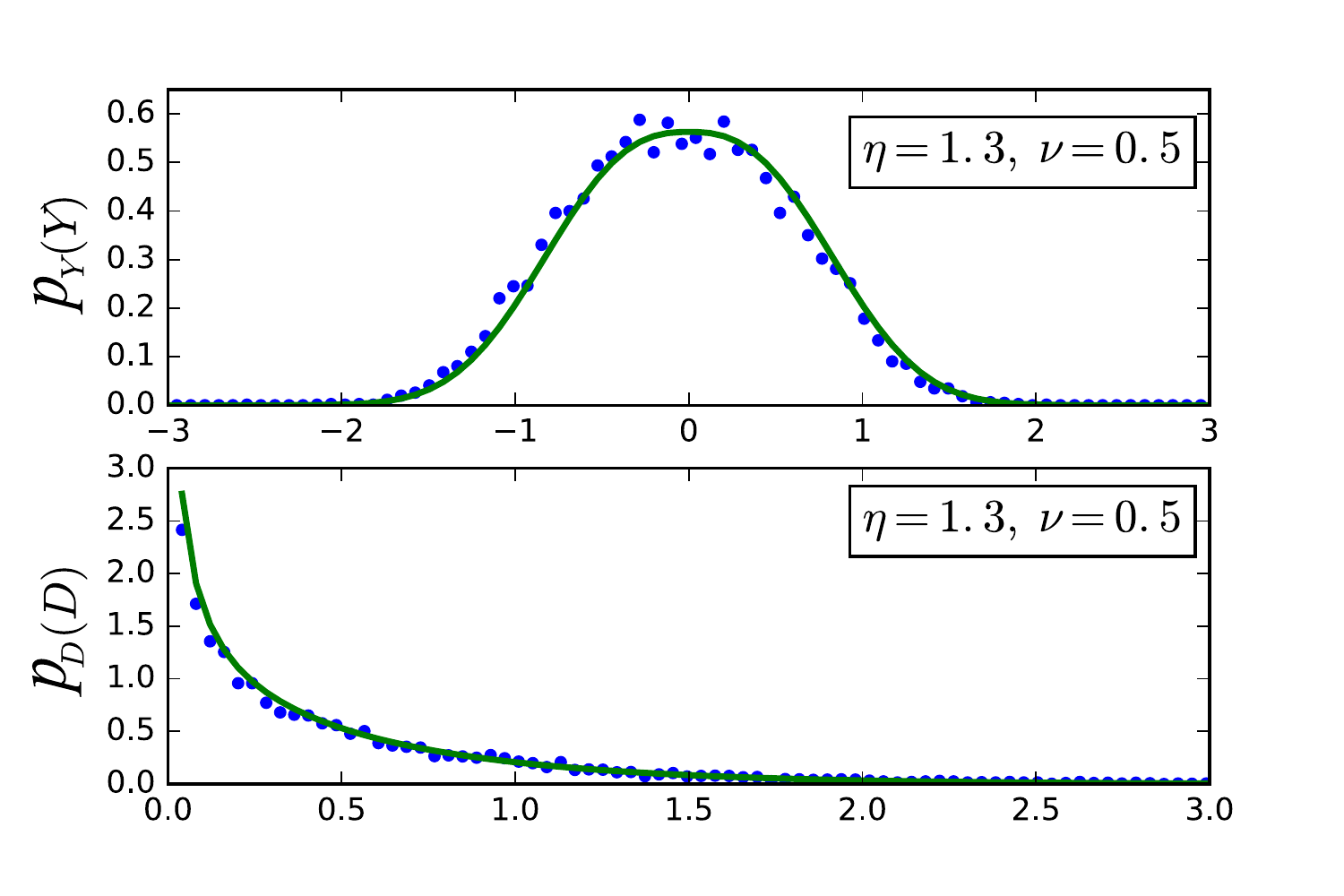}
\includegraphics[width=0.49\textwidth]{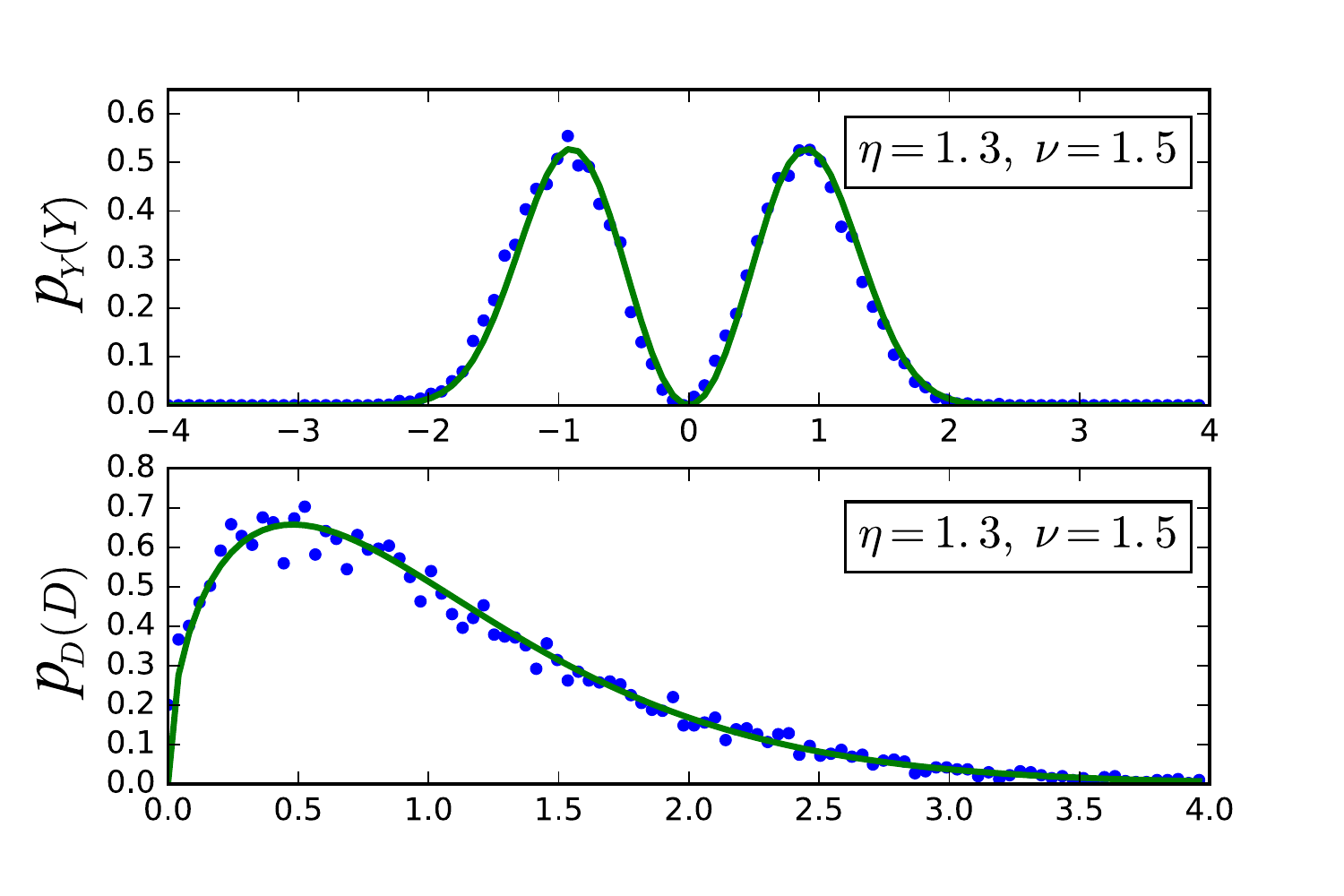}
\caption{PDFs of the auxiliary variable $Y(t)$ and the random diffusivity
$D(t)$ for two different sets of parameters, as indicated in the figure
legends.}
\label{img_rand_diff_PDF}
\end{figure}

Equation (\ref{diff_syst})
can be readily solved numerically with initial conditions taken randomly from
the equilibrium distribution (\ref{PDF_y}).
Figures \ref{img_rand_diff_1} and \ref{img_rand_diff_2} show sample time evolutions
of the auxiliary variable $Y$ and the diffusivity $D=Y^2$ for the DD process based
on the steady state generalised Gamma distribution, as obtained below. We note that
for the case $\nu=0.5$ in figure \ref{img_rand_diff_1} the sample paths of the
variable $Y(t)$ frequently cross the zero line, while for the case $\nu=1.5$ in
figure \ref{img_rand_diff_2} the zero line is avoided, corresponding to the uni-
and bi-modal shapes of the PDFs of the variable $Y(t)$ evaluated in figure
\ref{img_rand_diff_PDF}. The existence of a pole in the generalised Gamma
distribution (\ref{D_PDF}) at $D=0$ for the case $\nu=0.5$ thus creates a very
different behaviour than for the case $\nu=1.5$ without singularity. For the
diffusivity
variable $D(t)$ in figures \ref{img_rand_diff_1} and \ref{img_rand_diff_2} the
regions of $Y(t)$ close to the zero line lead to smaller $D(t)$ values in the
same regions. Finally, figures \ref{img_rand_diff_1} and \ref{img_rand_diff_2}
demonstrate the exponential shape of the autocorrelation functions for both
$Y(t)$ and $D(t)$,
\begin{equation}
\label{auto}
\mathrm{ACF}_Y(t,t')=\langle(Y(t)-\langle Y\rangle)(Y(t+t')-\langle Y\rangle)
\rangle
\end{equation}
and an analogous expression for $D(t)$.

We know from previous studies of DD models that the correlation time of the
random diffusivity represents a key factor in the study of the particle dynamics.
The correlation time $\tau_c$ is evaluated both by means of a two-parametric
numerical fit to the exponential function and through the integral
\begin{equation}
\tau_c\sim\frac{1}{\mathrm{ACF}(0)}\int_0^\infty\mathrm{ACF}(\tau)d\tau,
\end{equation}
which is exact for pure exponential autocorrelation functions. The results
obtained by the two methods are reported in figure \ref{img_rand_diff_1}
and \ref{img_rand_diff_2} and they are in excellent agreement, from which
we conclude that the diffusivity autocorrelation is exponential to leading
order and thus the correlation time $\tau_c$ well defined.

It is interesting to notice that the auxiliary function $Y(t)$ in the case
of a bimodal distribution possesses a non-zero correlation function in the
stationary state. This is due to the fact that despite a vanishing global
mean of the PDF, depending on the initial setting each trajectory is
representative of only one side of the bimodal PDF.

\section{A generalised minimal model for diffusing diffusivities}
\label{DD_sec}

\begin{figure}
\centering
\includegraphics[width=0.49\textwidth]{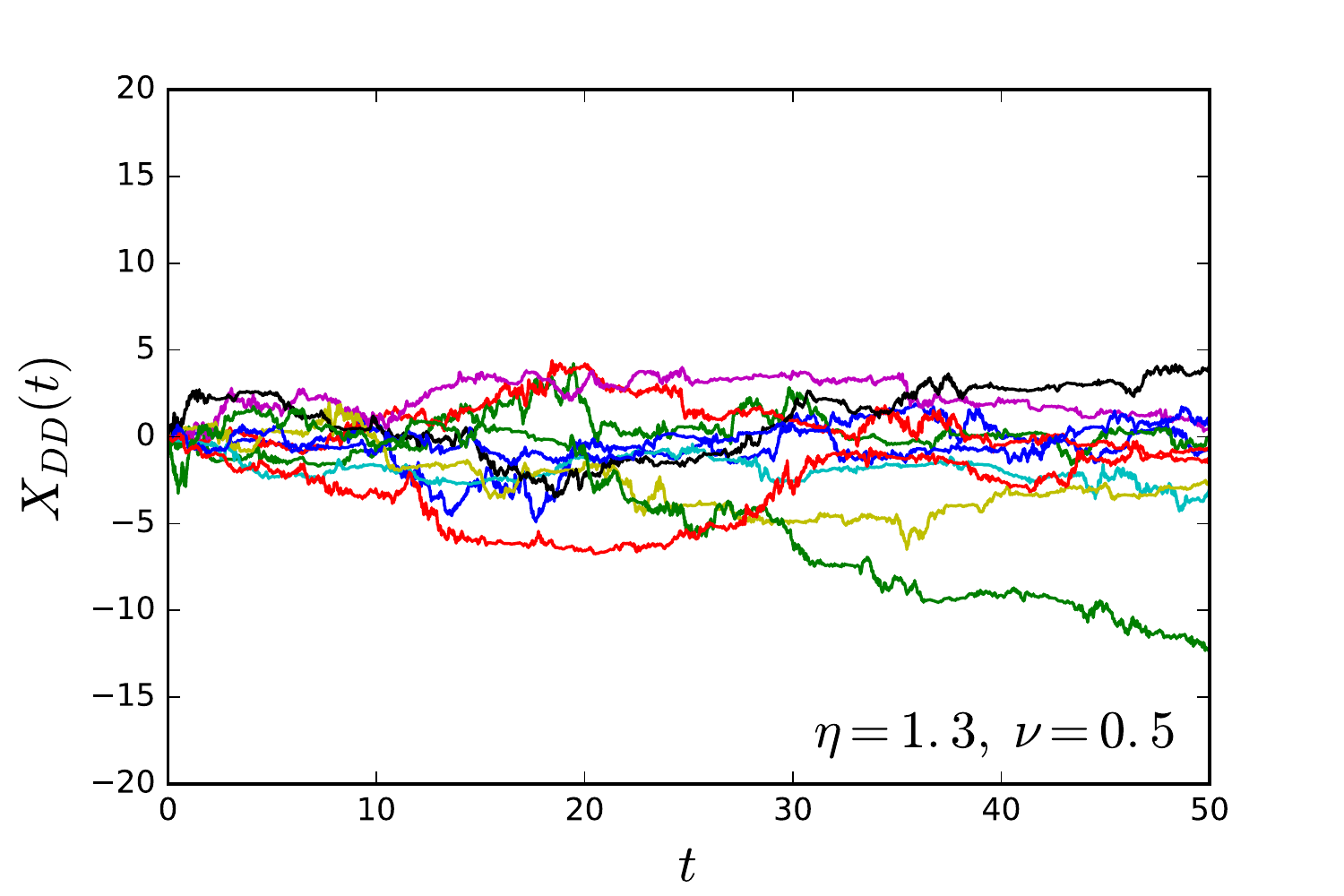}
\includegraphics[width=0.49\textwidth]{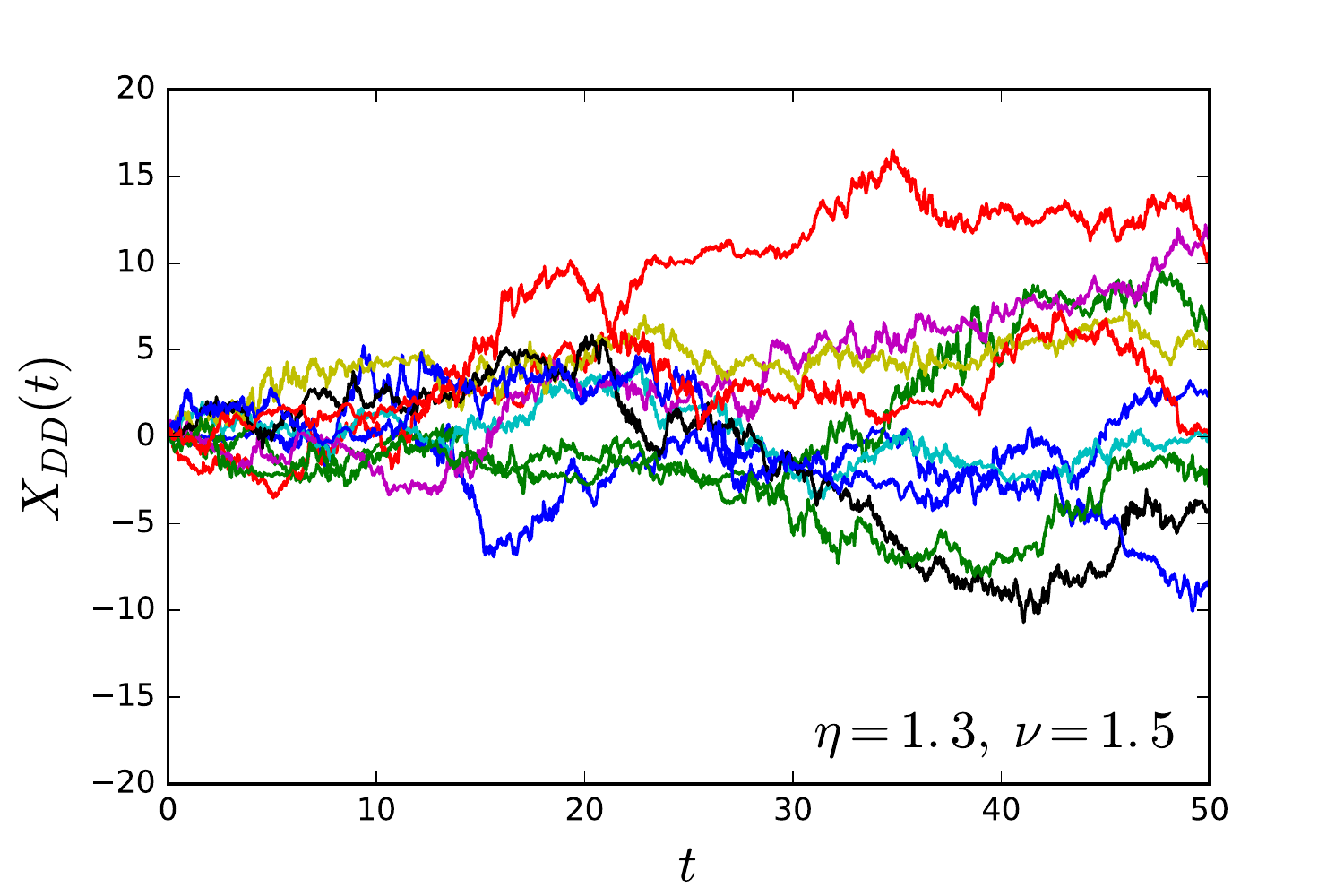}
\includegraphics[width=0.49\textwidth]{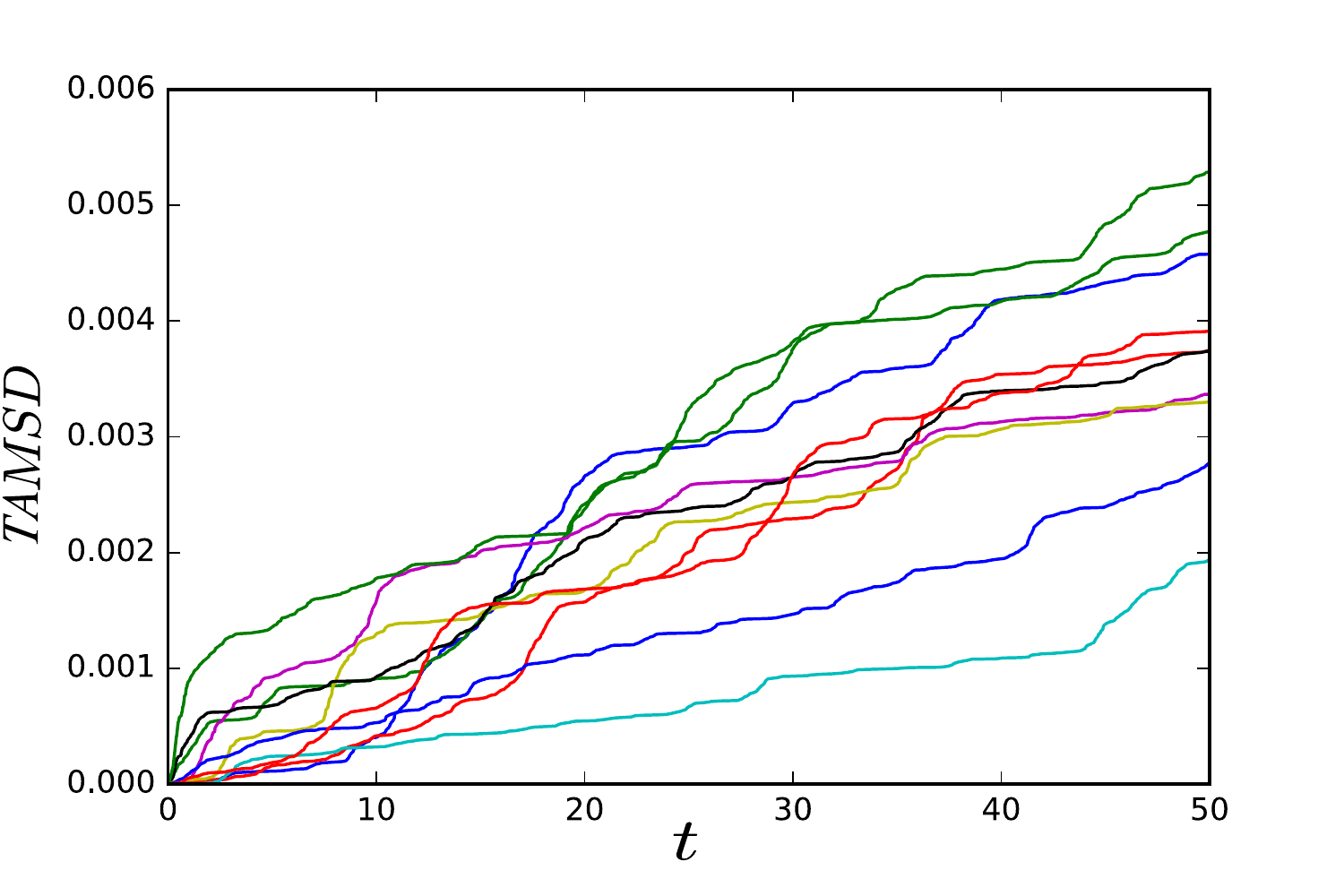}
\includegraphics[width=0.49\textwidth]{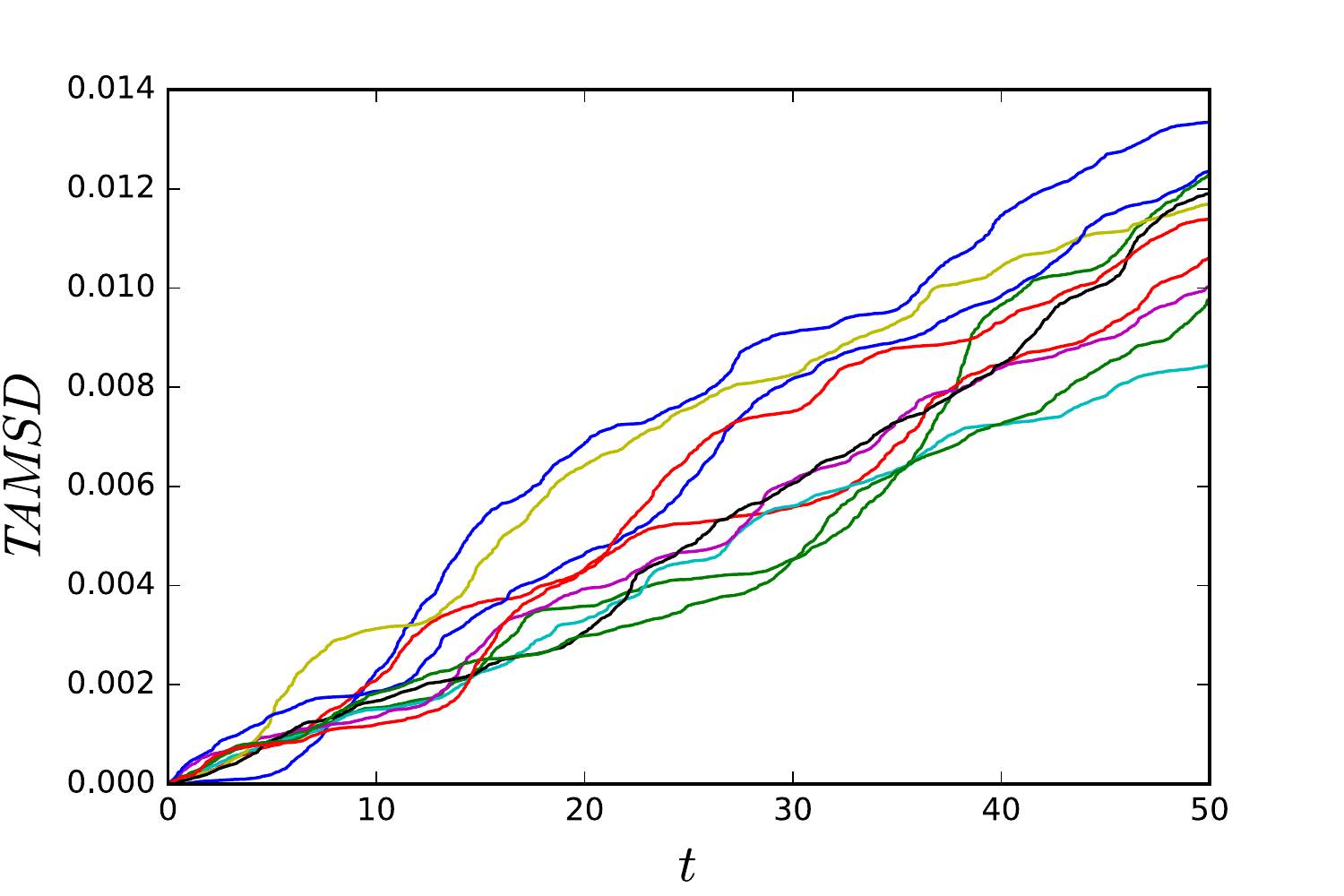}
\caption{Top: Trajectories of the DD model for two different sets of parameters
$\nu$ and $\eta$, as indicated in the figure legend and bottom: corresponding
time averaged MSDs. In contrast to the behaviour of the ggBM model shown in figure
\ref{traj_ggBM}, the temporal variation of the diffusivity $D(t)$ is distinct.}
\label{traj_DD}
\end{figure}

With the set of equations defined in section \ref{rand_diff} we can consider
the generalisation of the DD minimal model described in \cite{Chechkin:DD2},
and obtain the process in position space, $X_\mathrm{DD}(t)$. Recalling the
idea of introducing an analytic description for the dynamics of the random
diffusivity, we take that the motion of the particle is defined by the
integral version of the overdamped Langevin equation,
\begin{equation}
X_\mathrm{DD}(t)=\int_0^t\sqrt{2D(t')}\times\xi(t')dt',
\label{DD_x}
\end{equation}
where $ \xi(t)$ is white Gaussian noise and $D(t)$ is the random time dependent
diffusivity obtained in section \ref{rand_diff}. This dynamics based on the
above results for the diffusivity dynamics generalises the idea introduced in
\cite{Chechkin:DD2}, where an Ornstein-Uhlenbeck process was selected for the
auxiliary variable. Figure \ref{traj_DD} shows trajectories obtained from the
stochastic equation (\ref{DD_x}) where the diffusivity was generated from
(\ref{diff_syst}) with initial conditions taken randomly from the stationary
distribution. In ggBM
each trajectory has the same $D$ value, while in the DD model the value of $D$
changes as function of time. In turn, individual trajectories of the DD model
are quite similar.

Since the DD model is a direct generalisation of the minimal DD
model we expect a crossover to a Gaussian displacement PDF for times longer
than the correlation time $\tau_c$. We thus carry on our analysis for the
short and long time regimes separately, before analysing the MSD and kurtosis
of this DD process.

\subsection{Short time regime}
\label{ST}

Since the dynamics of the environment is determined by the correlation time
$\tau_c$ we expect that on short time scales with $t\ll\tau_c$ the diffusion
coefficient is approximately fixed for each particle and we thus suppose the
validity of a superstatistical description at short times (ST),
\begin{equation}
X_\mathrm{DD}^\mathrm{ST}\sim\sqrt{2D}\int_0^t\xi(t')dt'=\sqrt{2D}\times W(t).
\end{equation}
The existence of the superstatistical regime at $t\ll\tau_c$ is consistent with
the model considered in \cite{Chechkin:DD2} and with the results reported in
\cite{tyagi:DD5} concerning the modulation of white noise by any stochastic
process whose time correlation function decays exponentially. The superstatistical
approach allows us to estimate the short time distribution of the particle
displacement by means of
\begin{equation}
f^\mathrm{ST}_\mathrm{DD}(x,t)\sim\int_0^{\infty}p_D(D)G(x,t|D)dD.
\label{DD_PDF_ST}
\end{equation}
This representation corresponds to the ggBM scenario established above, which
means that we can borrow its results in equations (\ref{PDF_ST_Hfunc}) and
(\ref{asymp_PDF_ggBM_Hfunc}), considering that $f^\mathrm{ST}_\mathrm{DD}(x,t)
\sim f_\mathrm{ggBM}(x,t)$.

\begin{figure}
\centering
\includegraphics[width=0.49\textwidth]{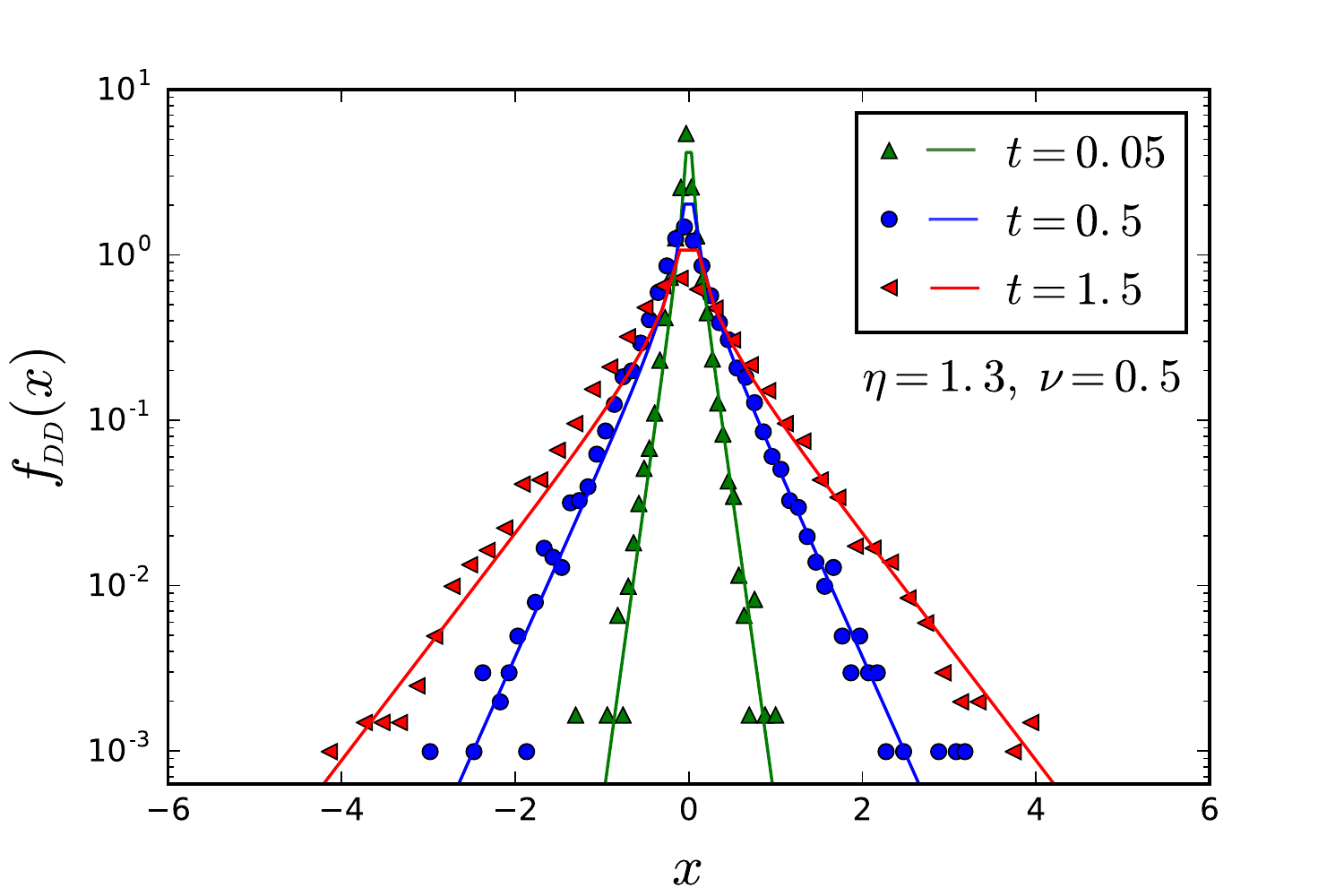}
\includegraphics[width=0.49\textwidth]{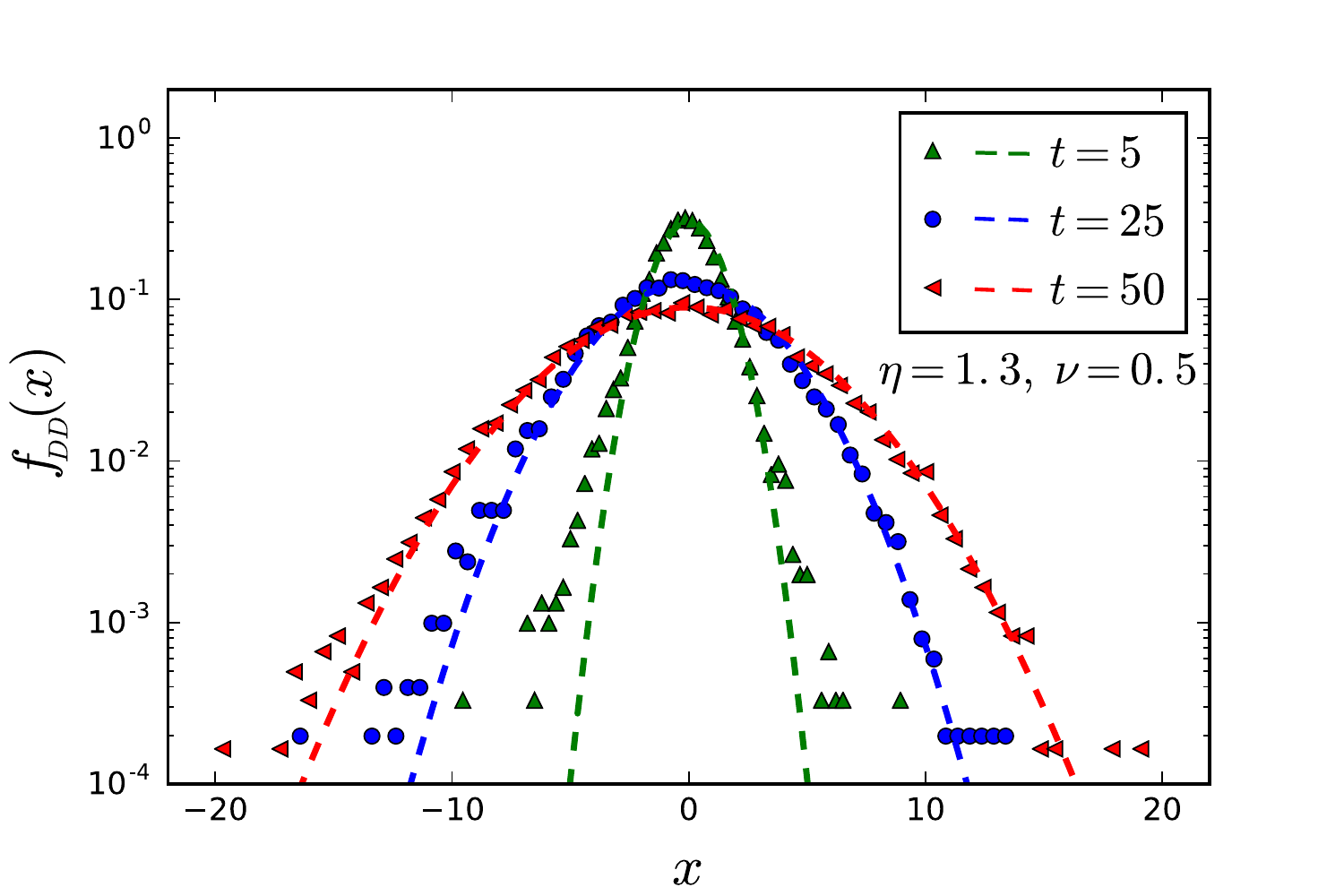}
\caption{Short time (a) and long time (b) PDF of the DD model for $\eta=1.3$
and $\nu=0.5$. The solid lines represent the asymptotic behaviour
(\ref{asymp_PDF_ggBM_Hfunc}) while the dashed lines represent the Gaussian
behaviour (\ref{PDF_DD_LT}) expected at sufficiently long times.}
\label{img_DD_1}
\end{figure}

The expected behaviour (\ref{asymp_PDF_ggBM_Hfunc}) is confirmed by extensive
numerical simulations. Figures \ref{img_DD_1}(a) and  \ref{img_DD_2}(a) show
the short time PDFs for two different sets of the parameters $\nu$ and $\eta$,
and in both cases we observe excellent agreement with the asymptotic behaviour
(\ref{asymp_PDF_ggBM_Hfunc}).

Comparing figure \ref{img_ggBM} with figure \ref{img_DD_1}(a) we notice that
the ggBM model allows one to describe a process that preserves the exact
non-Gaussian PDF, which is exactly the same PDF we obtain in the DD model in
the short time regime. Both approaches describe the same superstatistical
frame but the DD model then crosses over to a Gaussian beyond the correlation
time $\tau_c$, see below the discussion of the kurtosis. The establishment of
the relation between the DD model and the previously devised ggBM at short
times is our third main result.

\subsection{Long time regime}
\label{LT}

\begin{figure}
\centering
\includegraphics[width=0.49\textwidth]{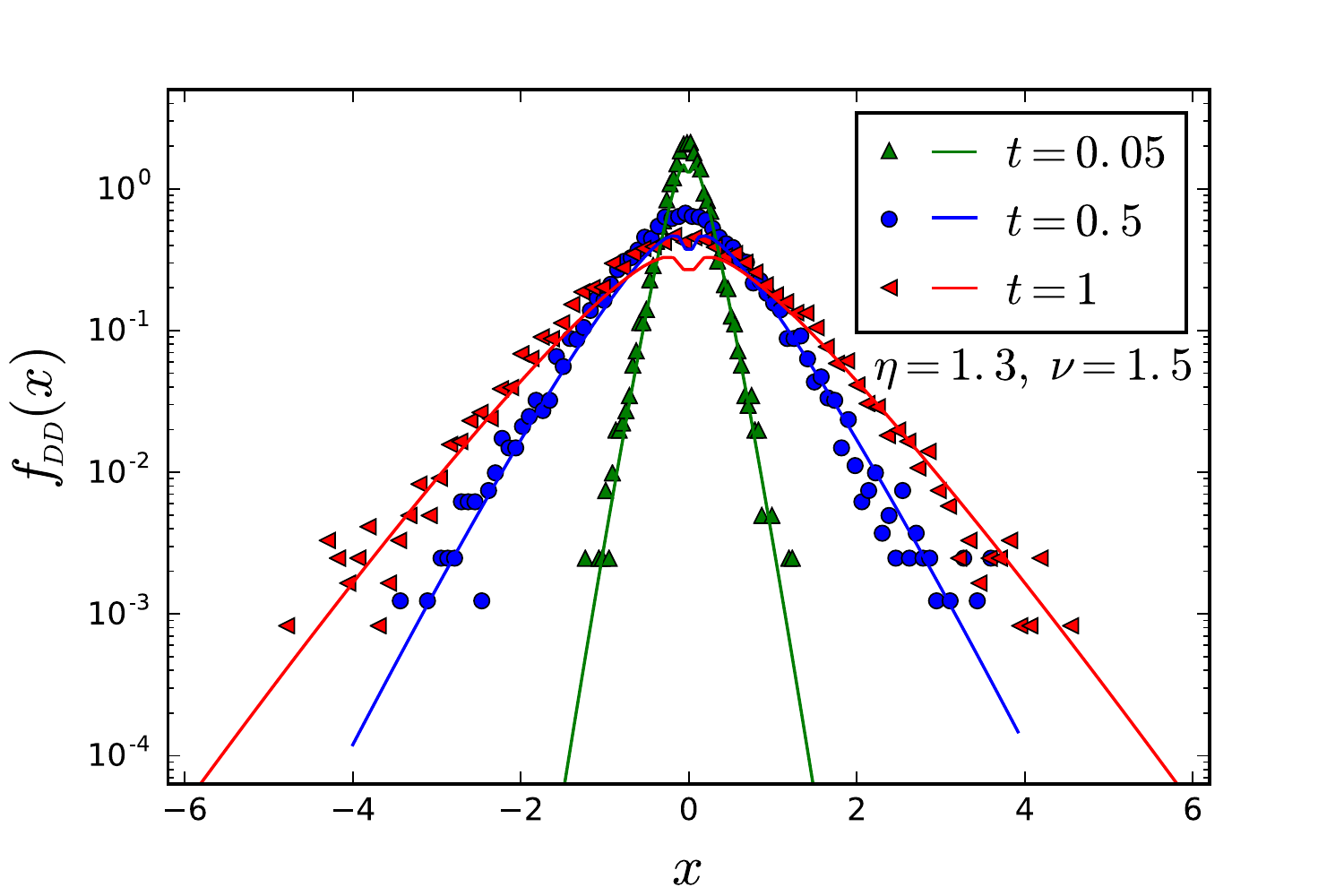}
\includegraphics[width=0.49\textwidth]{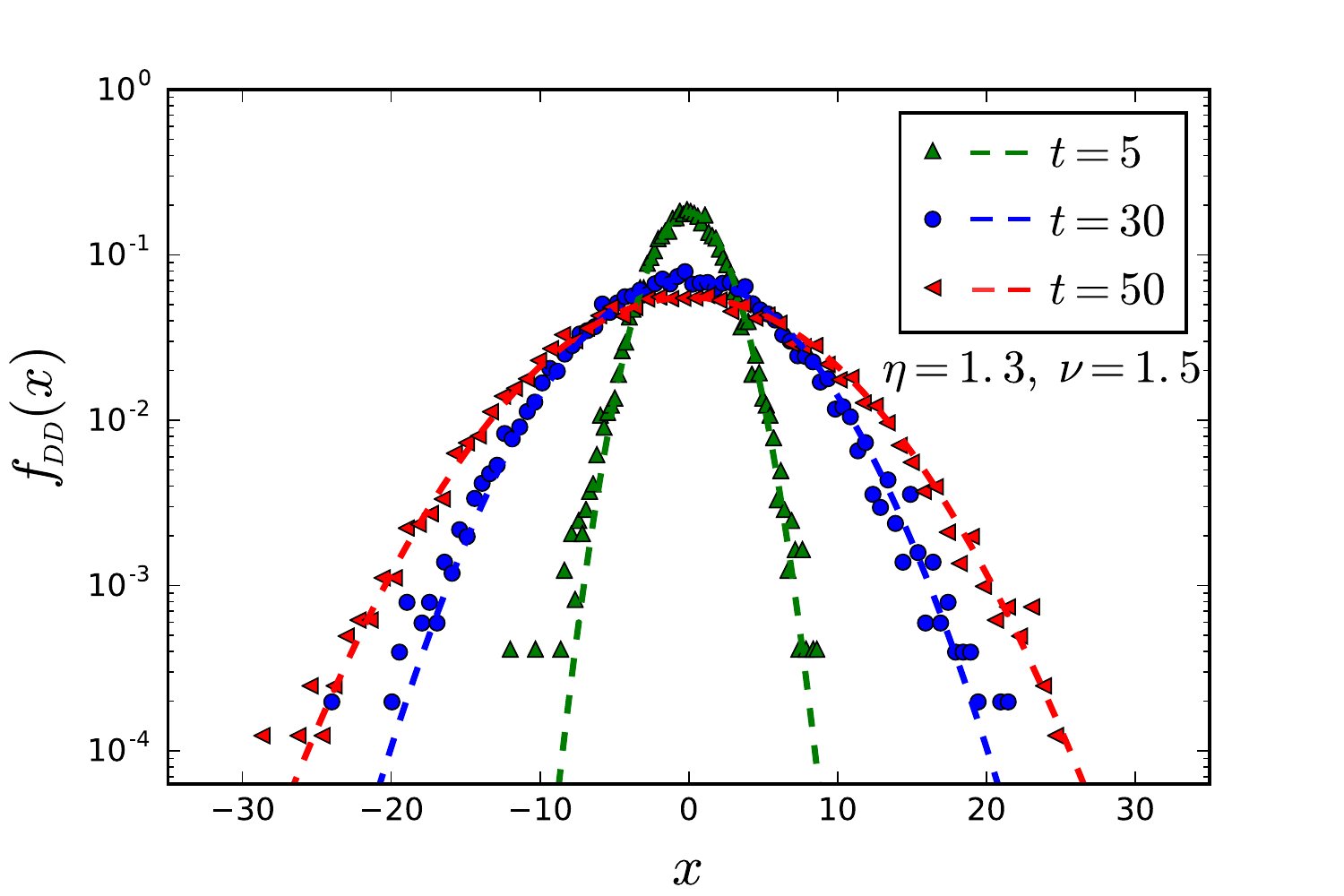}
\caption{Short time PDF (a) and long time PDF (b) of the DD model for $\eta=1.3$
and $\nu=1.5$. The solid lines represent the asymptotic behaviour in
(\ref{asymp_PDF_ggBM_Hfunc}) while the dashed lines represent the Gaussian
behaviour in (\ref{PDF_DD_LT}) at long times.}
\label{img_DD_2}
\end{figure}

At long times (LT), again taking our clue from \cite{Chechkin:DD2} and from the
general results in \cite{tyagi:DD5}, we expect that eventually a crossover
to a Gaussian distribution is observed (as already anticipated in figures
\ref{img_DD_1} and \ref{img_DD_2}). Above the correlation time, that is, for
times $t\gg\tau_c$ we thus look for a PDF given by
\begin{equation}
f^\mathrm{LT}_\mathrm{DD}(x,t)\sim\frac{1}{\sqrt{4\pi\langle D\rangle_\mathrm{
stat}t}}\exp\left(-\frac{x^2}{4\langle D\rangle_\mathrm{stat}t} \right),
\label{PDF_DD_LT}
\end{equation}
with the effective diffusivity (\ref{d_eff}). The numerical results reported in
figures \ref{img_DD_1}(b) and  \ref{img_DD_2}(b) prove the validity of this
behaviour. At sufficient long times the particles have explored all
the diffusivity space and a Gaussian behaviour with an effective diffusivity
emerges. This leads to a standard Brownian diffusive behaviour. We stress again
that the transition from a non-Gaussian to a Gaussian profile depends on
the value of the correlation time $\tau_c$ of the diffusivity process.

\subsection{Mean squared displacement}

\begin{figure}
\centering
\includegraphics[width=0.41\textwidth]{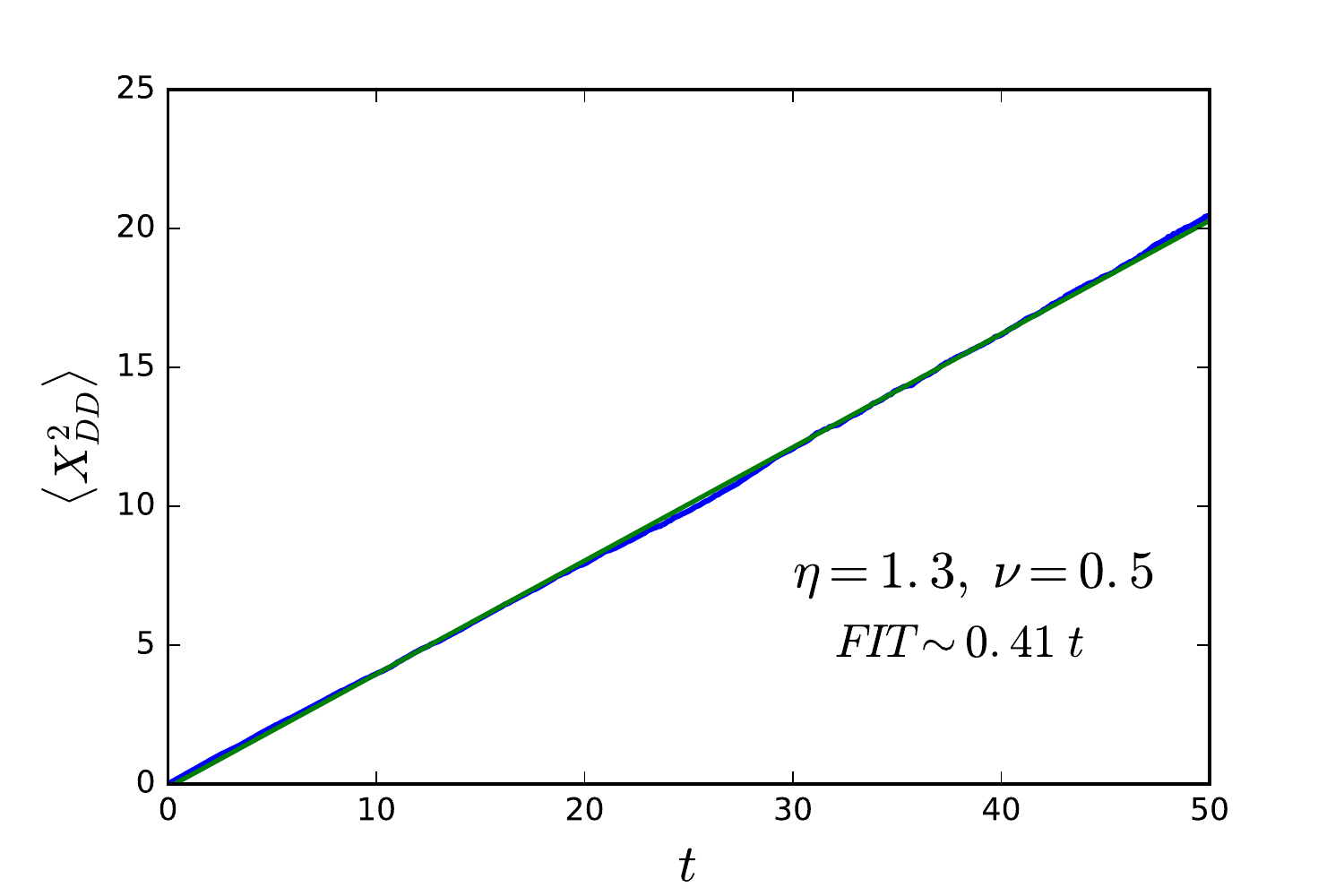}
\caption{Variance of the DD model. The solid green line represents a linear
fit and the corresponding slope is reported in the plot. It is consistent
with the expected value $0.40$ according to equation (\ref{d_eff}).}
\label{img1_moments}
\end{figure}

For the DD model we found a crossover of the PDF of the spreading particles. An
initial non-Gaussian behaviour is slowly replaced by a Gaussian one. The
superstatistical behaviour of the DD approach at short times is equivalent to
the ggBM model and is characterised by the non-Gaussianity. Nevertheless, as
expected from previous studies \cite{Chechkin:DD2}, the MSD does not change in
the course of time and is the same at short and long time regimes. Direct
calculation indeed produces the invariant form
\begin{equation}
\label{ddmsd}
\langle X_\mathrm{DD}^2(t)\rangle=2\langle D\rangle_\mathrm{stat}t.
\end{equation}
This continuity of the MSD is demonstrated in figure \ref{img1_moments},
together with a linear fit proving the validity of the linear trend.

\subsection{Kurtosis}
\label{kurtosis}

\begin{figure}
\centering
\includegraphics[width=0.49\textwidth]{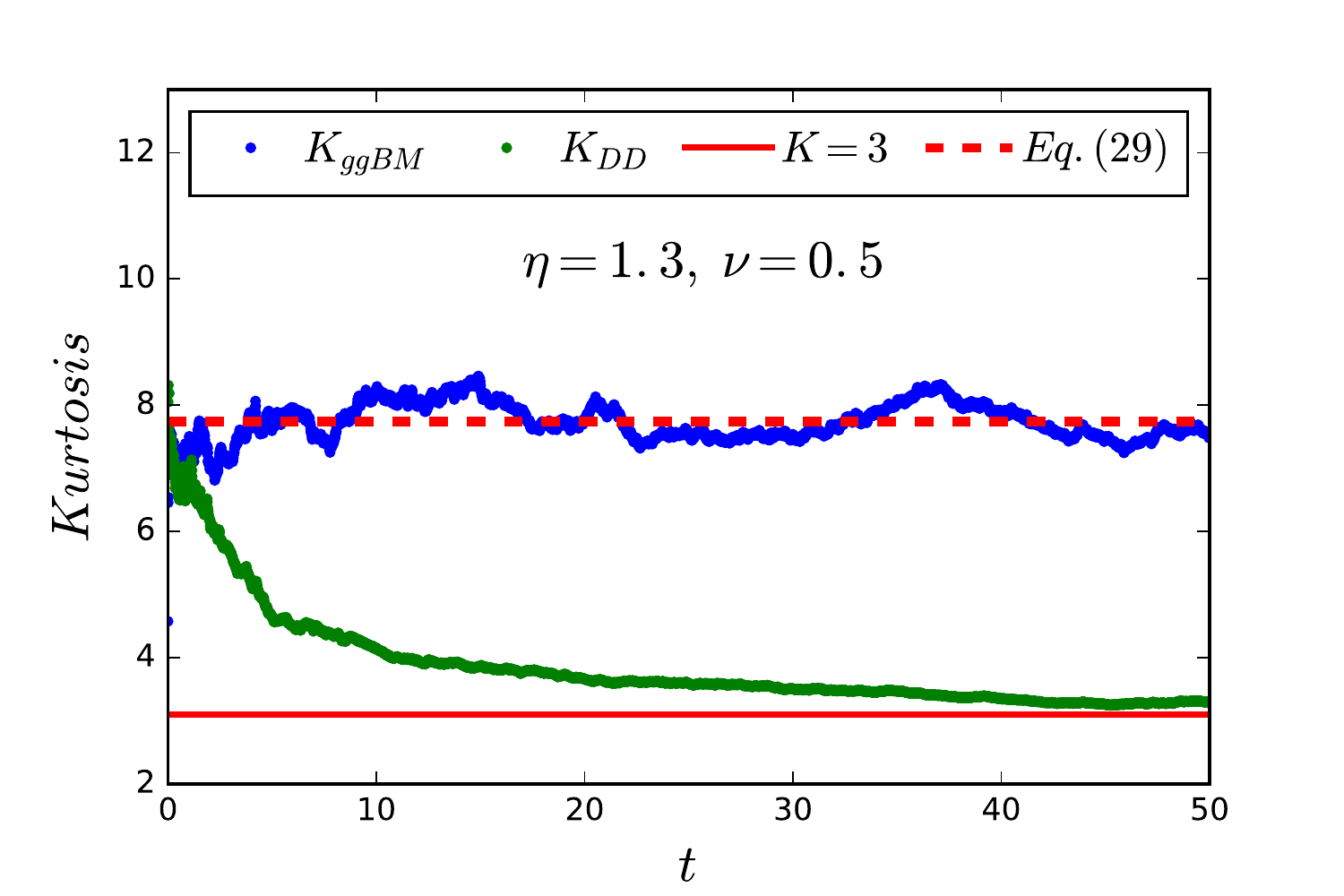}
\caption{Kurtosis of the DD process (green) and ggBM (blue) for 10,000
realisations. For the indicated value of $\eta$ and $\nu$ equation
(\ref{K}) yields $K_\mathrm{ggBM}\approx7.74$.}
\label{img2_moments}
\end{figure}

In what follows the second and fourth moments of the non-Gaussian PDF identified
in equations (\ref{ggBM_PDF}) and (\ref{DD_PDF_ST}) are studied in terms of the
kurtosis that represents one of the first checks for non-Gaussianity. We recall
the second order moment calculated in (\ref{2_mom}) and in a similar way we
obtain the fourth order moment
\begin{equation}
\langle X_\mathrm{ggBM}^4(t)\rangle=\langle X_\mathrm{DD}^4(t)\rangle_\mathrm{ST}
=12\langle D^2\rangle_{stat}t^2,
\label{4_mom}
\end{equation}
where $\langle D^2\rangle_{stat}$ is the second moment of the diffusivity in
the stationary state. By means of results (\ref{2_mom}) and (\ref{4_mom}) and
recalling the definition of the diffusivity moments in equation (\ref{moments}),
the kurtosis $K=\langle x^4(t)\rangle/\langle x^2(t)\rangle^2$ is given by
\begin{eqnarray}
K_\mathrm{ggBM}&=&K_\mathrm{DD}^\mathrm{ST}=4t^2D_*^2\frac{3\Gamma([\nu+2]/\eta)}{
\Gamma(\nu/\eta)}\times\frac{1}{4t^2D_*^2}\left(\frac{\Gamma(\nu/\eta)}{\Gamma
([\nu+1]/\eta)}\right)^2\nonumber\\
&=&3\frac{\Gamma([\nu+2]/\eta)\Gamma(\nu/\eta)}{\Gamma([\nu+1]/\eta)^2}>3,
\label{K}
\end{eqnarray}
for ggBM and the short-time DD process. The non-Gaussian PDF represents a
leptokurtic behaviour as can be observed in figure \ref{img2_moments},
showing the kurtosis of the DD and ggBM models. The value for the kurtosis
at short times is in agreement with the value reported in (\ref{K}). At long
times the DD kurtosis approaches the value 3 characteristic of the Gaussian
distribution, while the ggBM one keeps fluctuating around the same initial
value.

\section{Non-equilibrium initial conditions}
\label{NEIC}

The results discussed above consider equilibrium initial conditions for the
diffusivity fluctuations. In particular, results (\ref{2_mom}) and (\ref{ddmsd})
for the particle MSD exhibit the invariant form $\langle X^2(t)\rangle=2\langle
D\rangle_{\mathrm{stat}}t$ in both cases. Such equilibrium initial conditions
will in general not be fulfilled for particles that are initially seeded in a
non-equilibrium environment. For instance, in single particle tracking a tracer bead
can be introduced into the system at $t=0$, or similar in computer simulations.
After this disturbance the environment equilibrates again. To accommodate for such
a case we here study a minimal model for the case of non-equilibrium initial
conditions, which leads to another main result of this work. As we are going to
see, this non-equilibrium scenario gives rise to differences in the characteristics
of the two studied models. In particular, we observe an initial ballistic behaviour.
The long time behaviour, of course, does not show differences since in this range
the diffusivity reaches its stationary state and we can again consider the results
obtained in the previous sections for the long time limit.

We illustrate the role of non-equilibrium conditions by taking a specific, and
in fact the simplest, set of parameters, $\nu=0.5$ and $\eta=1$. This defines the
stochastic dynamical equation in (\ref{diff_syst}) as
\numparts
\begin{eqnarray}
\label{OU_NEIC_1}
D(t)&=Y^2(t)\\
dY&=-\frac{\sigma^2}{D_\star}Ydt+\sigma dW(t),
\label{OU_NEIC_2}
\end{eqnarray}
\endnumparts
that corresponds to the well known dynamics of the Ornstein-Uhlenbeck process
for the study in \cite{Chechkin:DD2} with the correlation time $\tau_c=D_\star/
\sigma^2$. We start considering the related Fokker-Planck equation
\begin{equation}
\frac{\partial}{\partial t}p(Y,t)=\frac{\sigma^2}{D_\star}\frac{\partial}{\partial
Y}Yp(Y,t)+\frac{\sigma^2}{2}\frac{\partial^2}{\partial Y^2}p(Y,t).
\end{equation}
We can solve this equation with a non equilibrium condition, for instance,
$p(Y,0)= \delta(Y-Y_0)$, using the method of characteristics in Fourier space.
We readily derive the general solution
\begin{equation}
\fl
p(Y,t|Y_0)=\left(\pi D_\star[1-\exp(-2t\sigma^2/D_\star)]\right)^{-1/2}\exp\left(
-\frac{(Y- Y_0\exp(-t\sigma^2/D_\star))^2}{D_\star(1-\exp(-2t\sigma^2/D_\star))}
\right).
\label{y_NEIC}
\end{equation}
Recalling relation (\ref{PDF_rel}) for the diffusivity PDF we then obtain
\begin{eqnarray}
\nonumber
p_D(D,t|D_0=Y_0^2)&=&\frac{1}{2\sqrt{D}}\left[p(\sqrt{D},t)+p(-\sqrt{D},t)\right]\\
\nonumber
&=&(4\pi D_\star D[1-e^(-2t\sigma^2/D_\star)])^{-1/2}\\
&\times&\left\{\exp\left(-\frac{(\sqrt{D}-\sqrt{D_0}\exp(-t\sigma^2/D_\star))^2}{
D_\star(1-\exp(-2t\sigma^2/D_\star))}\right)\right.\nonumber\\
&&\left.+\exp\left(-\frac{(-\sqrt{D}-\sqrt{D_0}\exp(-t\sigma^2/D_\star))^2}{
D_\star(1-\exp(-2t\sigma^2/D_\star))}\right)\right\}.
\label{D_NEIC}
\end{eqnarray}
We point out that in the limit of long times this result provides exactly
the stationary distribution described in (\ref{D_PDF}) with the specific
set of parameters defined above. This is also verified by the trend of the
average value
\begin{equation}
\langle D(t)\rangle=\frac{1}{2}\left(D_\star(1-e^{-2t\sigma^2/D_\star})+2D_0
e^{-2t\sigma^2/D_\star}\right),
\label{mean_D}
\end{equation}
in agreement with result (\ref{D_PDF}).

In contrast to the previous analysis, we observe an explicit dependence on
time of $p_D(D,t)$, which makes the calculations more involved. Thus, we
select an initial condition for the diffusivity, $D_0=0$, which is convenient
for the study of the particles displacement distribution.  This leads to a
reduction in (\ref{D_NEIC}), namely,
\begin{equation}
\fl
p(D,t|D_0=0)=(\pi D_\star D(1-\exp(-2t\sigma^2/D_\star)))^{-1/2}
\exp\left(\frac{D}{D_\star(1-\exp(-2t\sigma^2/D_\star))}\right).
\label{D_NEIC_D0}
\end{equation}
We now study the two models in this particular case of a non-equilibrium
initial condition for the diffusivity.

\subsection{Diffusing diffusivities with non-equilibrium initial diffusivity
condition}

The dynamics for the diffusivity encoded in equations (\ref{OU_NEIC_1})
and (\ref{OU_NEIC_2}) when choosing the specific set of parameters $\nu=0.5$
and $\eta=1$ is the same as described in \cite{Chechkin:DD2} when $d=n=1$. Thus,
in this paragraph, we extend the description of the minimal DD model studied
in \cite{Chechkin:DD2} to the case of non-equilibrium initial conditions for
the diffusivity. In order to proceed with the same notation we introduce
dimensionless units for relations (\ref{OU_NEIC_1}) and (\ref{OU_NEIC_2}) as
well as for the overdamped Langevin equation describing the particle motion
\cite{Chechkin:DD2}, such that the full set of stochastic equations reads
\begin{eqnarray}
X_\mathrm{DD}&=\int_0^t\sqrt{2D(t')}\xi(t')dt'\nonumber\\
D(t)&=Y^2(t)\nonumber\\
dY&=-Ydt+dW(t).
\label{DD_NEIC}
\end{eqnarray}
A subordination approach can then be used to obtain the distribution of the
particle displacement \cite{Chechkin:DD2}, namely,
\begin{equation}
f_{DD}(x,t)=\int_0^\infty T(\tau,t)G(x,\tau)d\tau, 
\label{DD_sub}
\end{equation}
where $G(x,\tau)$ is the Gaussian (\ref{gauss}) and $T(\tau,t)$ represents the
probability density function of the process $\tau(t)=\int_0^t Y^2(t')dt'$.
Starting from the subordination formula (\ref{DD_sub}) we obtain the relation
\begin{equation}
\hat{f}_{DD}(k,t)=\tilde{T}(s=k^2,t)
\label{f_DD_NEIC}
\end{equation}
where with the symbols $\hat{\cdot}$ and $\tilde{\cdot}$ we indicate the Fourier
and Laplace transforms, respectively. For the particular initial condition
$D_0=0$, which is equivalent to $y_0=0$, the solution is known \cite{Lipster,dankel},
\begin{equation}
\fl\tilde{T}(s,t)=\exp\left(\frac{t}{2}\right)\left/\left(\frac{1}{\sqrt{1+2s}}
\sinh(t\sqrt{1+2s})+\cosh(t\sqrt{1+2s})\right)^{1/2}\right..
\label{T_NEIC}
\end{equation}
This latter quantity is directly related to the MSD of the particles through
\cite{Chechkin:DD2}
\begin{equation}
\langle X^2_\mathrm{DD}(t)\rangle=-2\left.\frac{\partial\tilde{T}(s,t)}{\partial
s}\right|_{s=0}. 
\end{equation}
We readily obtain the closed form result
\begin{equation}
\langle X^2_\mathrm{DD}(t)\rangle=t-\frac{1}{2}(1-e^{-2t})\sim\left\{
\begin{array}{ll} t^2, & t\ll1\\ t, & t\gg1\end{array}\right..
\label{var_DD_NEIC}
\end{equation}
The resulting dynamics is thus no longer Brownian at all times. In contrast,
at times shorter than the correlation time (in the dimensionless units used
here $\tau_c=1$) we obtain a ballistic scaling of the MSD. This behaviour
reflects the fact that the diffusivity equilibration in this case with $D_0=0$
leads to an initial acceleration.

Starting from equations (\ref{f_DD_NEIC}) and (\ref{T_NEIC}) we consider
approximations of the PDF for short and long times which, since we are in
dimensionless units, correspond to $t\ll1$ and $t\gg1$ respectively. In the
short time limit, the Fourier transform of the PDF becomes
\begin{equation}
\hat{f}^{ST}_{DD}(k,t)\sim\frac{(1+t+t^2/2)^{1/2}}{(1+t+t^2/2+k^2t^2)^{1/2}}
\sim\frac{1}{t}\left(k^2+\frac{1}{t^2}\right)^{-1/2}.
\end{equation}
Note that this expression is normalised, $\hat{f}_{DD}(k=0,t)=1$. After
taking the inverse Fourier transform we find
\begin{eqnarray}
\nonumber
f^{ST}_{DD}(x,t)&\sim&\frac{1}{\pi t}\int_0^\infty\frac{\cos(kx)}{(k^2+1/t^2)^{
1/2}}dk\\
&=&\frac{1}{\pi t}K_0\left(\frac{|x|}{t}\right).
\end{eqnarray}
Re-establishing dimensional units, this result becomes
\begin{equation}
f^{ST}_{DD}(x,t)\sim\frac{1}{\pi\sigma t}K_0\left(\frac{|x|}{\sigma t}\right).
\end{equation}
Here $K_\nu(z)$ is the modified Bessel function of second type. The asymptotic
behaviour of this distribution for $|x|\to\infty$ is the Laplace distribution
\begin{equation}
f^{ST}_{DD}(x,t)\sim\frac{1}{\sqrt{2\pi\sigma t|x|}}\exp\left(-\frac{|x|}{\sigma
t}\right).
\label{NEIC_DD_ST}
\end{equation}

In the long time limit equations (\ref{f_DD_NEIC}) and (\ref{T_NEIC}) yield
\begin{equation}
\hat{f}^{LT}_{DD}(k,t)\sim\frac{2^{1/2}\exp(t[1-\sqrt{1+2k^2}]/2)}{(1+1/\sqrt{
1+2k^2})^{1/2}} ,
\end{equation}
that again is normalised. If we focus on the tails of the distribution in the
limit $k\ll1$ we obtain the Gaussian
\begin{equation}
\hat{f}^{LT}_{DD}(k,t)\sim\exp(-k^2 t/2)
\end{equation}
in Fourier space, corresponding to the Gaussian
\begin{equation}
f^{LT}_{DD}(x,t)\sim\frac{1}{\sqrt{2\pi t}}\exp\left(-\frac{x^2}{2t}\right)
\end{equation}
in direct space. Restoring dimensional units and recalling that $\langle D\rangle
_\mathrm{stat}=D_\star/2$, eventually provides
\begin{eqnarray}
f^{LT}_{DD}(x,t)&\sim&\frac{1}{\sqrt{2\pi D_\star t}}\exp\left(-\frac{x^2}{2D_\star
t}\right)\nonumber\\
&=&\frac{1}{\sqrt{4\pi\langle D\rangle_\mathrm{stat}t}}\exp\left(-\frac{x^2}{4
\langle D\rangle_\mathrm{stat}t}\right),
\label{NEIC_DD_LT}
\end{eqnarray}
where in the last step we identified the equilibrium value $\langle D\rangle
_\mathrm{stat}$ of the diffusivity. From the approximations (\ref{NEIC_DD_ST})
and (\ref{NEIC_DD_LT}) we readily recover the two limiting scaling laws for the
variance in equation (\ref{var_DD_NEIC}).

\begin{figure}
\centering
\includegraphics[width=0.49\textwidth]{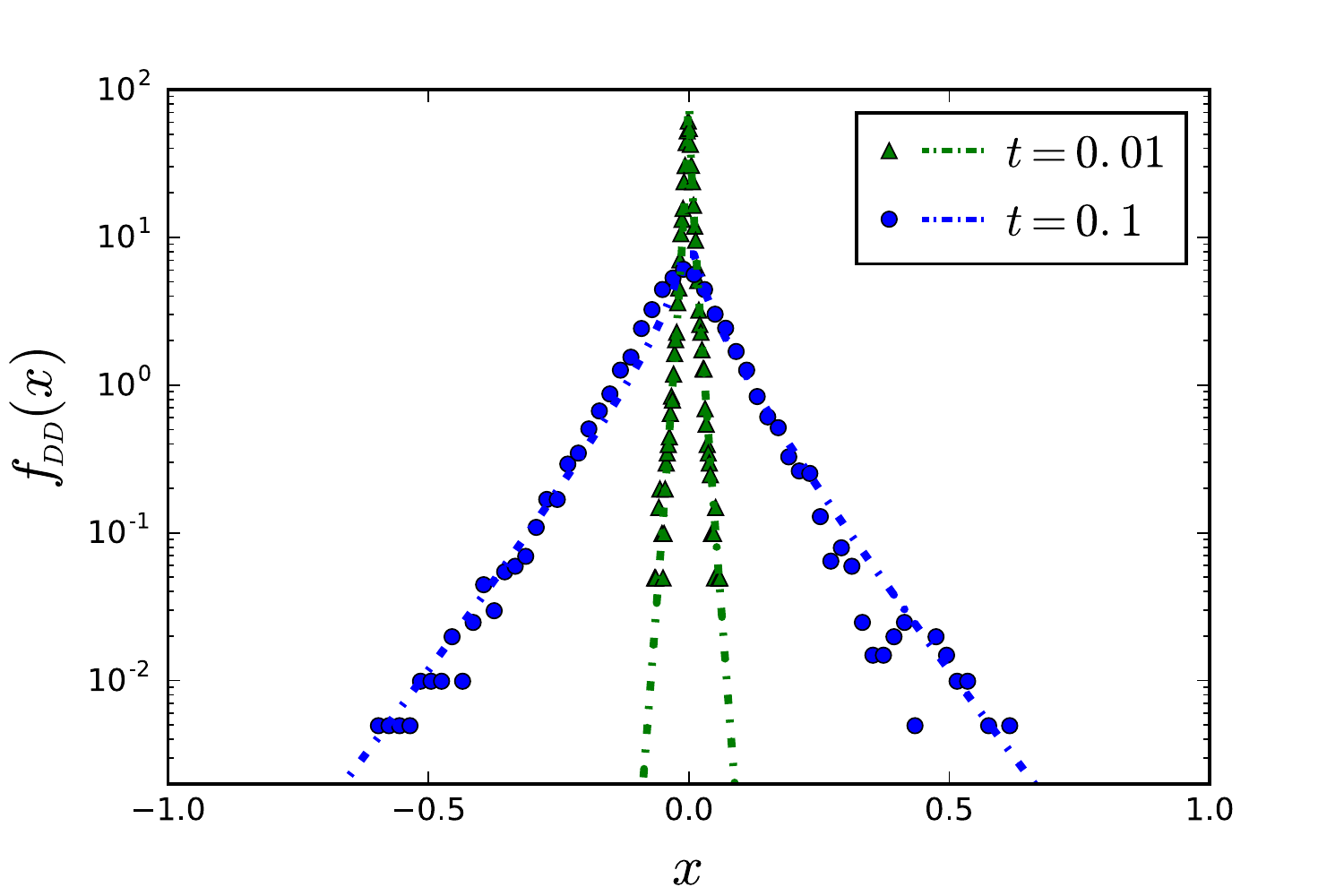}
\includegraphics[width=0.49\textwidth]{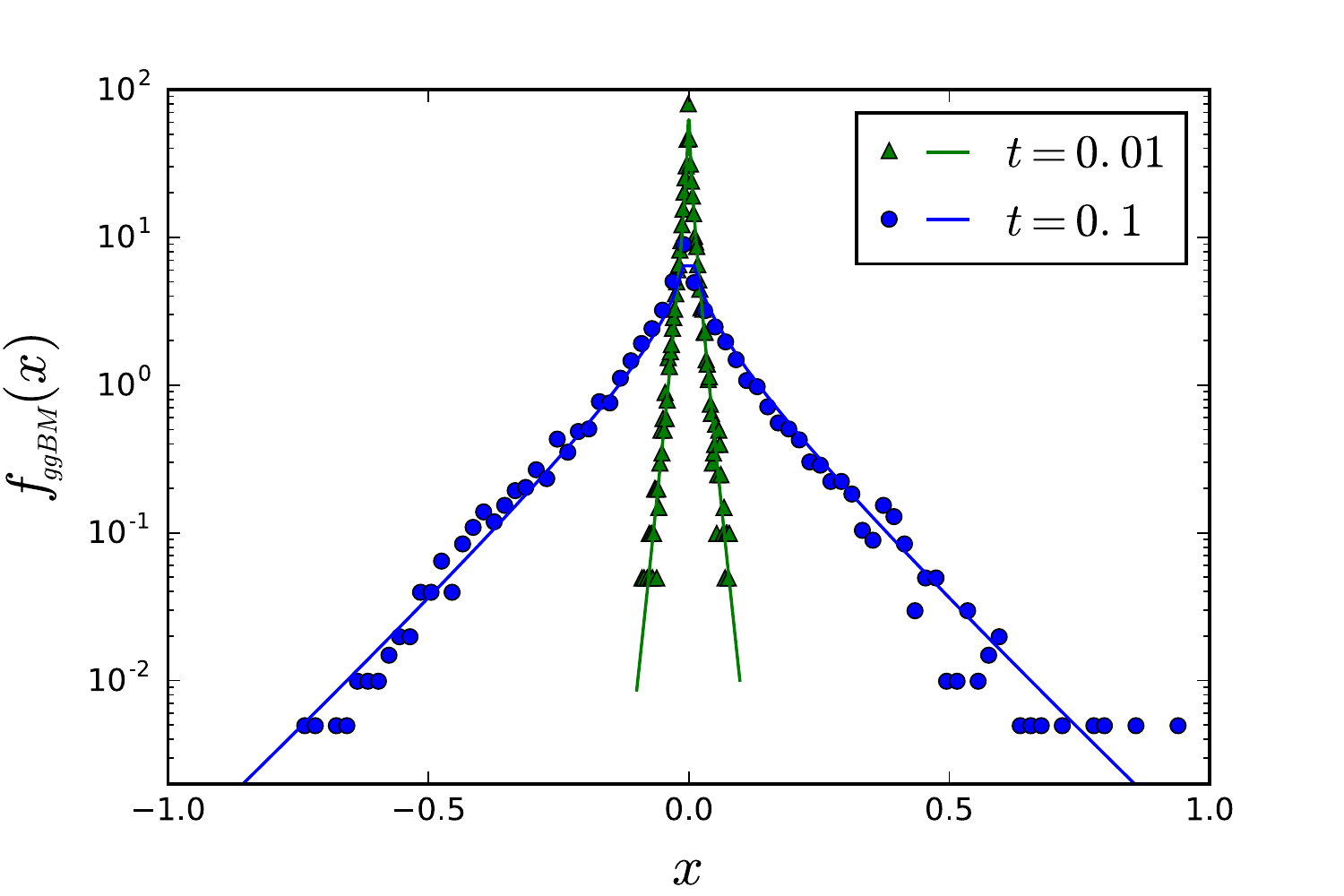}\\
\includegraphics[width=0.49\textwidth]{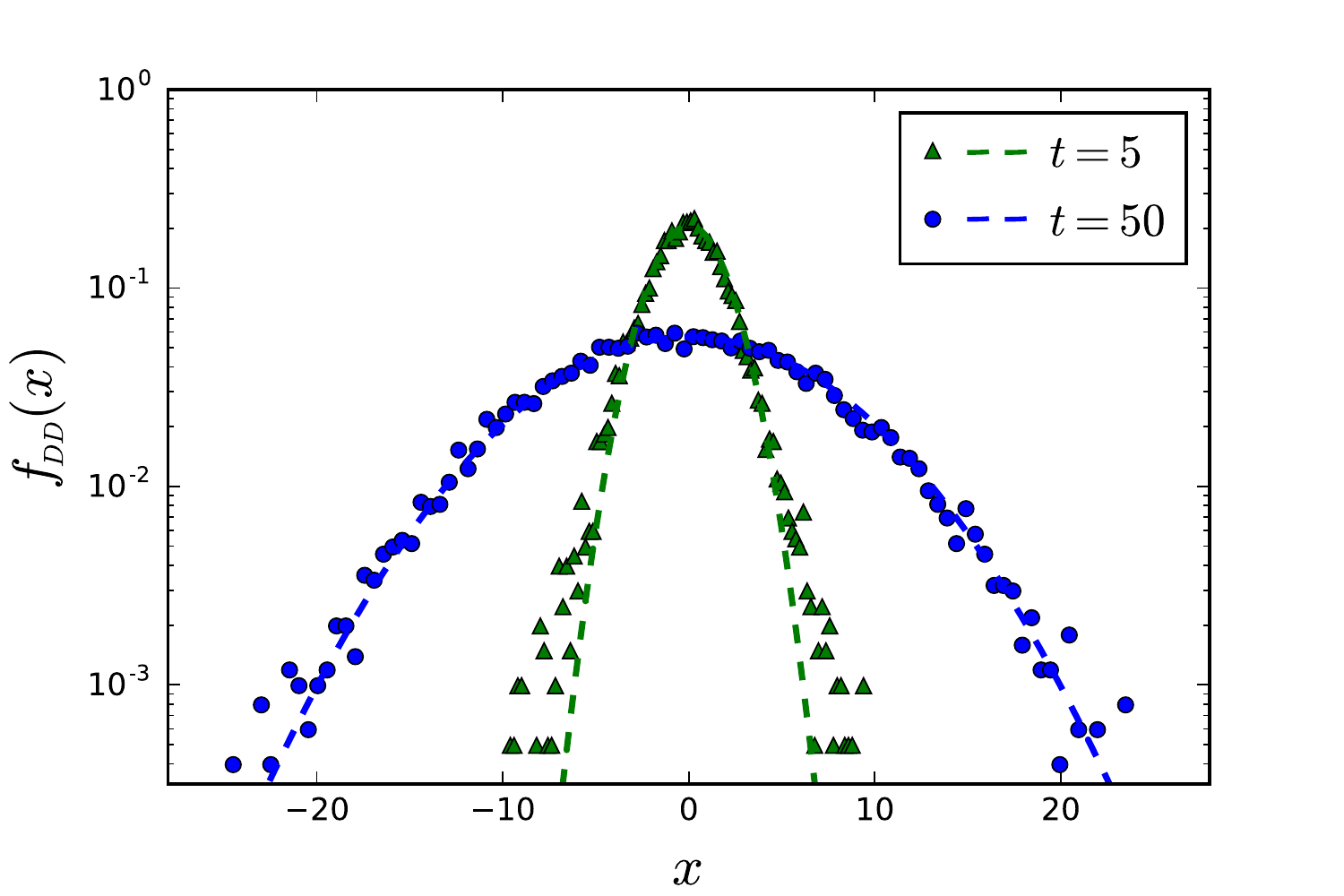}
\includegraphics[width=0.49\textwidth]{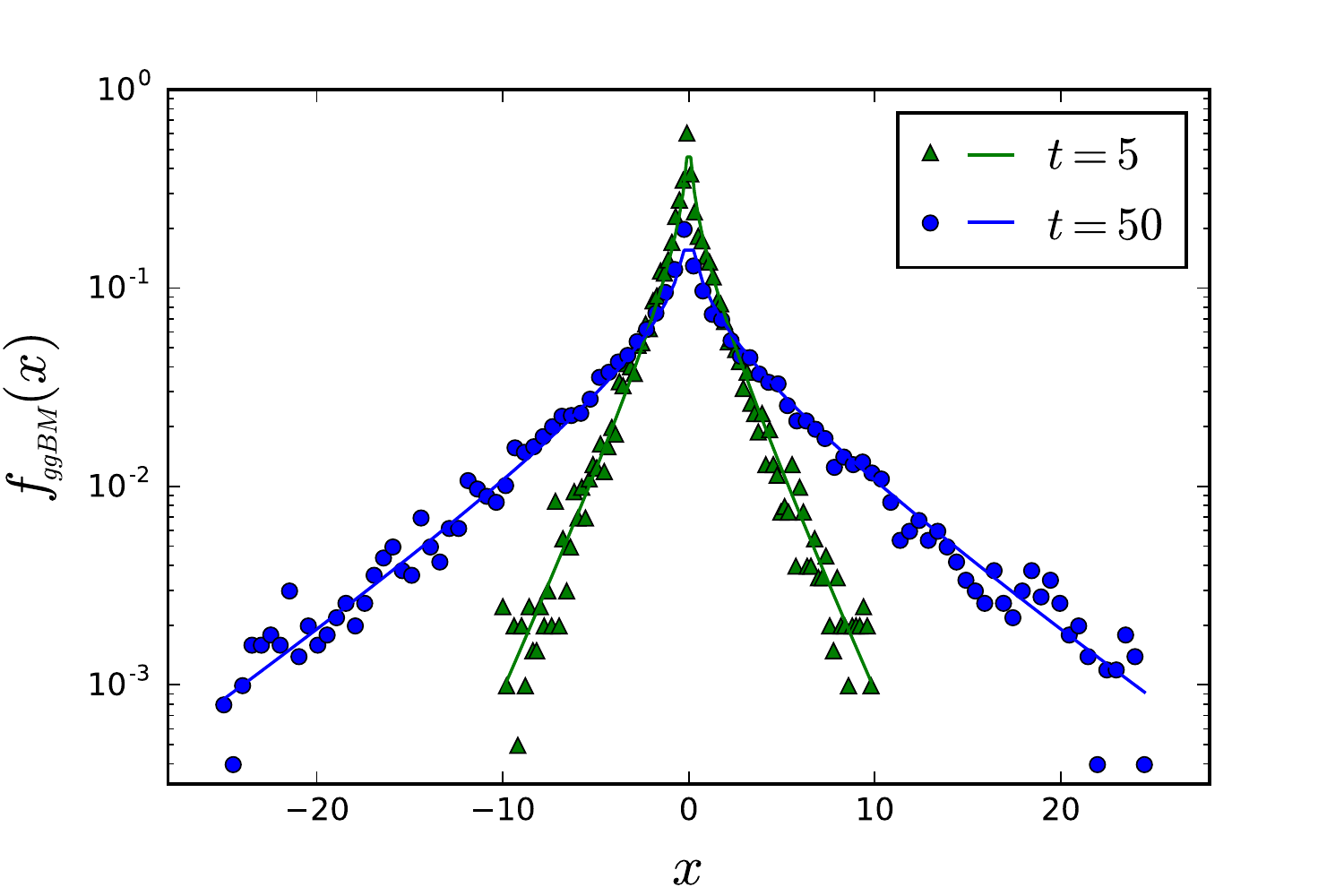}
\caption{PDFs of the DD (left) and ggBM (right) models with non-equilibrium
initial condition $D_0=0$ of the diffusivity. Top: short time behaviour. Bottom:
long time behaviour. For the DD model, the dash-dotted lines represent the
asymptotic behaviour (\ref{NEIC_DD_ST}) at short times, while the dashed lines
are Gaussian fits. For the ggBM model the solid lines represent the analytical
result (\ref{NEIC_ggBM}).}
\label{NEIC_pdfs}
\end{figure}

\begin{figure}
\centering
\includegraphics[width=0.49\textwidth]{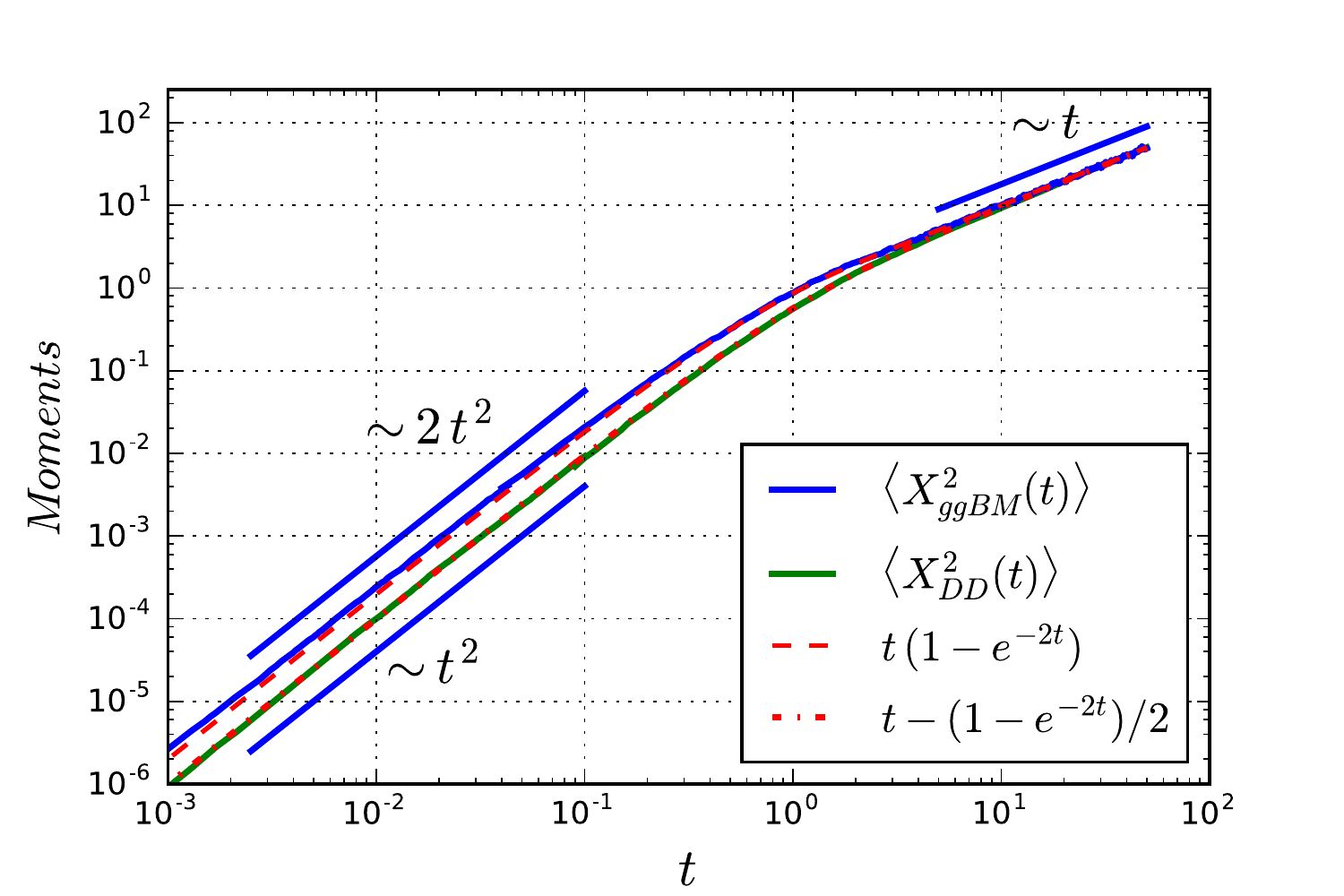}
\includegraphics[width=0.49\textwidth]{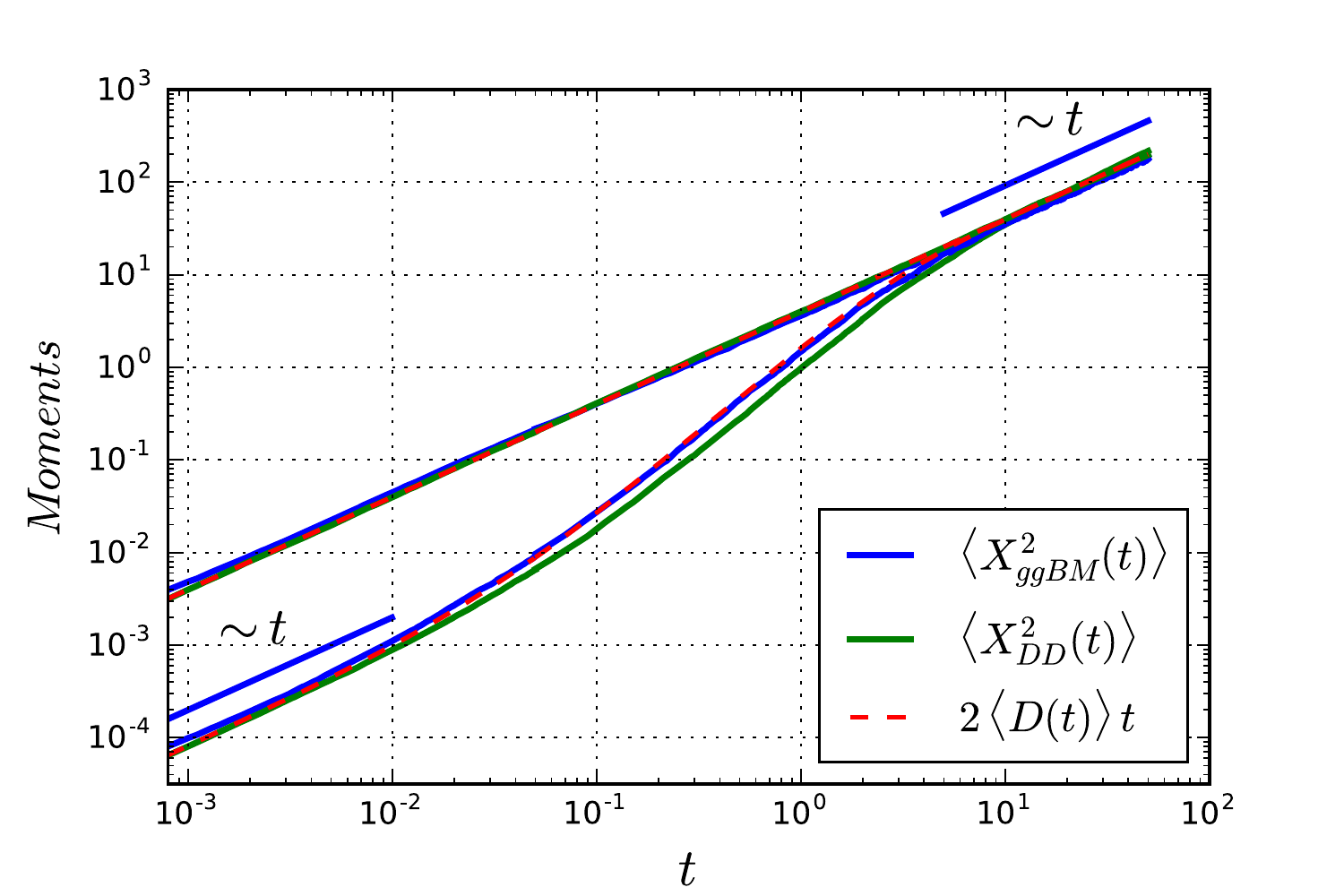}
\caption{MSD of the DD model (green) and ggBM (blue). On the left $D_\star=1$
and $D_0=0$ while on the right we have $D_\star=4$ and two different values of
$D_0$, $D_0=\langle D\rangle_\mathrm{stat}=D_\star/2$ and $D_0=0.04$. The first
value generates a linear trend of the variance for both models, as we saw for
the equilibrium case. In the second case, where $D_0\ll D_\star$, we observe
three regimes. Nice agreement with the analytical results is observed.}
\label{NEIC_moments}
\end{figure}

Figure \ref{NEIC_pdfs} nicely corroborates these findings, comparing the
non-equilibrium DD model results for the PDF obtained above with results
from stochastic simulations. The crossover behaviour of the associated MSD
is displayed in figure \ref{NEIC_moments}, again showing very good agreement
with the theory.

\subsection{Non-equilibrium generalised grey Brownian motion}

The ggBM model discussed in section \ref{ggBM_sec} is based on the static
distribution $p_D(D)$ of the diffusivity. In order to explore non-equilibrium
effects as discussed above for the DD model also within the superstatistical
approach, we here propose a non-equilibrium generalisation of the ggBM model.
Thus, we generalise the standard ggBM definition (\ref{ggBM_x}) and introduce a
variability of $D$ in time, according to the stochastic equation
\begin{equation}
\label{ggBM_x_t}
X_\mathrm{ggBM}(t)=\sqrt{2D(t)}\times W(t).
\end{equation}
Physically, this new concept may be interpreted as fluctuations of the disjointed
environments experienced by the different particles or to temporal changes of the
particle size, for instance, due to agglomeration-separation dynamics.

Based on the definition (\ref{ggBM_x_t}) it is then straightforward to take the
dynamics of $D(t)$ to be the same as the one considered for the DD model. This
guarantees that the ensemble properties of this generalised process (\ref{ggBM_x_t})
are exactly the same as the ones of the standard ggBM model studied in section
\ref{ggBM_sec}. In particular, the dependence on time of the diffusivity does not
affect the validity of equation (\ref{lemma_2}), so in order to
estimate the PDF of the particle displacement of the ggBM model,
we consider the distribution (\ref{D_NEIC_D0}) in the calculation of the integral
\begin{equation}
f_\mathrm{ggBM}(x,t|D_0)=\int_0^{\infty}p_D(D,t|D_0)G(x,t|D)dD,
\label{dyn_super}
\end{equation}
which may be defined in general as a dynamic superstatistics because of the
dependence of $p_D(D,t)$ on $t$. We obtain an explicit solution by means of
the Mellin transform following the same procedure as described in Appendix
\ref{app_2},
\begin{eqnarray}
\nonumber
f_\mathrm{ggBM}(x,t|D_0=0)&=&(\pi^2D_\star t(1-\exp(-2t\sigma^2/D_\star)))^
{-1/2}\\
&&\times K_0\left(\frac{|x|}{\sqrt{D_\star t(1-\exp(-2t\sigma^2/D_\star))}}\right),
\label{NEIC_ggBM}
\end{eqnarray}
where $K_\nu(z)$ is the modified Bessel function of second type. The asymptotic
behaviour for $|x|\to\infty$ is given by the exponential
\begin{equation}
\fl f_\mathrm{ggBM}(x,t|D_0=0)\sim\frac{1}{\sqrt{2\pi|x|\sqrt{D_\star t(1-e^{-2
t\sigma^2/D_\star})}}}\exp\left(-\frac{|x|}{\sqrt{D_\star t(1-e^{-2t\sigma^2/
D_\star})}}\right).
\end{equation}
However, in comparison with the result (\ref{asymp_PDF_ggBM_Hfunc}) in the
equilibrium situation we now observe a different time scaling in the exponent.
For short times we see that
\begin{equation}
f_\mathrm{ggBM}^{ST}(x,t|D_0=0)\sim\frac{1}{\sqrt{\sqrt{2}\pi\sigma t|x|}}
\exp\left(-\frac{|x|}{\sqrt{2}\sigma t}\right),
\label{NEIC_ggBM_ST}
\end{equation}
while at long times
\begin{equation}
f_\mathrm{ggBM}^{LT}(x,t|D_0=0)\sim\frac{1}{\sqrt{2\pi|x|\sqrt{D_\star t}}}
\exp\left(-\frac{|x|}{\sqrt{D_\star t}}\right).
\label{NEIC_ggBM_LT}
\end{equation}
Comparing the short time PDF in (\ref{NEIC_ggBM_ST}) with the DD model obtained
in (\ref{NEIC_DD_ST}) we notice that they show a difference in the time scaling
of a factor $\sqrt{2}$ which is exactly what we observe in figure \ref{NEIC_pdfs}.

Starting from equation (\ref{dyn_super}) the MSD can be written as
\begin{eqnarray}
\langle X_\mathrm{ggBM}^2(t)\rangle&=\langle\big(\sqrt{2D(t)}W(t)\big)^2\rangle
=2\langle D(t)\rangle t\nonumber\\
&=\left[D_\star(1-e^{-2t\sigma^2/D_\star})+2D_0e^{-2t\sigma^2/D_\star}\right]t.
\end{eqnarray}
Note that this result is valid for any initial conditions $D_0$, not only for the
case $D_0=0$. As already suggested above, the scaling of the variance is no longer
linear at all times. According to the relation between the parameters it is
possible to observe the different scaling behaviours
\begin{eqnarray}
\langle X_\mathrm{ggBM}^2(t)\rangle=\left\{\begin{array}{lll}2D_0t, & \sigma^2 t
\ll D_0\\
2D_0t+2\sigma^2t^2, & D_0\ll\sigma^2t\ll D_\star\\
D_\star t & \sigma^2t\gg D_\star\end{array}\right., & 0\le D_0\ll D_\star,\\
\langle X_\mathrm{ggBM}^2(t)\rangle=\left\{\begin{array}{ll}2D_0t-4\sigma^2
(D_0/D_\star)t^2 & \sigma^2t\ll D_\star\\
D_\star t & \sigma^2t\gg D_\star\end{array}\right. & \quad\mathrm{elsewhere.} 
\end{eqnarray}
Thus when $D_0\ll D_\star$ we observe three regimes for the MSD. When $D_0=0$
or when the relation $D_0\ll D_\star$ does not hold we directly observe an
initial ballistic behaviour followed by the stationary linear trend. This
behaviour is nicely corroborated in figure \ref{NEIC_moments}.

\section{Conclusions}
\label{concl}

A growing range of systems is being revealed which exhibit Brownian yet
non-Gaussian diffusion dynamics. Often, an exponential (Laplace) shape of
the displacement PDF is observed, however, also stretched Gaussian shapes
have been reported. The comparison of diffusion processes recorded by new
experimental techniques suggests that the complexity and inhomogeneity of the
medium, interpreted as the cause of non-Gaussian behaviour, may influence
the spreading of particles in specific fashion and at different levels. In
particular, experiments have demonstrated that a non-Gaussian dynamic may
persist throughout the observation window and that there are systems that,
instead, at long times, exhibit a crossover to Gaussian diffusion. In
this article we introduced an analytic approach to generate a random
and time-dependent diffusivity with specific features and we proposed two
possible models for the spreading dynamics of particles in complex systems:
one belonging to the class of generalised grey Brownian motion (ggBM) and
the other supporting the idea of diffusing diffusivities (DD).

We saw that the two models have in common the idea that the non-Gaussianity
of the PDF is a direct consequence of an inhomogeneity of the environment,
represented by a population of diffusivities. The same PDF for the random
diffusivity was introduced for both models. We defined an
operative set of dynamic stochastic equations to study random diffusivity
effects within the broad class of generalised Gamma distributions. This
includes the Gamma distribution, or the exponential PDF which produces
the Laplace distribution for the particle displacements.

We observed that the main difference between the ggBM and the DD model is the
description of the particle dynamics in the long time regime, corresponding
to different physical scenarios for the environment. GgBM does not consider
an active dynamics of the environment, and the characteristic that mainly
influences the particle motion is the randomness of the medium. This means
that the statistical features of the medium completely drive the particles in
their entire motion. In contrast the DD model supports the idea of diffusing
diffusivities considering a dynamics also for the environment. In this way
the particles evolve experiencing both a continuous variability in time
and a stochasticity in the ensemble. The first model delineates a specific
non-Gaussian dynamics for the entire diffusion process, while the second allows
to describe a transition from a non-Gaussian to a Gaussian diffusion. In fact,
it was shown that the short time non-Gaussian dynamics is the same in the two
models, whereas at longer times the ggBM model retains the diffusivity
distribution and the DD model leads to an effective value for the diffusivity.

We here also studied the influence of non-equilibrium initial conditions for the
diffusivity dynamics and found two main effects. First, the non-equilibrium
case breaks the equivalence of the DD and the ggBM models at short times and,
second, it causes changes in the temporal evolution of the MSD. In this case
the ggBM model, which we showed to represent a stochastic interpretation of
superstatistical Brownian motion, describes what we may call a dynamical
superstatistics that leads to the presence of different time scaling regimes
in the process. The DD model, which we investigated in this case via a
subordination approach, at short times can no longer be described through a
superstatistic approximation, since the subordination results in that regime
diverge from the behaviour of ggBM. Furthermore we observed different time
scaling regimes for the DD model, as well. Nevertheless, we note that for both
models we never obtained an anomalous time scaling for the MSD, only a
crossover between ballistic and linear (Brownian or Fickean) behaviour. In
the long time regime we obtained a description of the two models which is
in agreement with the one for the equilibrium case, as it should be.

It will be interesting to generalise the present findings to anomalous dynamics
with stochastic diffusivity by implementing different types of noise. Maintaining
the same population of diffusivities the results obtained for the PDF of the
particle displacement will not be affected, yet the MSD scaling will become
anomalous.

\ack

AVC and RM acknowledge funding from the DFG within project ME 1535/6-1.
This research is supported by the Basque Government through the BERC 2014-2017
programme and by the Spanish Ministry of Economy and Competitiveness MINECO
through BCAM Severo Ochoa excellence accreditation SEV-2013-0323 and through
project MTM2016-76016-R MIP.

\appendix

\section{Computation of the superstatistical integral}
\renewcommand\thesection{\Alph{section}}

In this Appendix we provide different methods to solve the integral representing
the non-Gaussian PDF of the two models discussed in this work,
\begin{equation}
\bar{P}(x,t)=\int_0^{\infty}p_D(D) G(x,t|D)dD,
\label{both_PDF}
\end{equation}
where $G(x,t|D)$ represents a Gaussian distribution and $p_D(D)$ is the
generalised Gamma distribution (\ref{D_PDF}).

\subsection{Computation via Fox $H$-function}
\label{app_1}
\label{app_1_Hfunc}

Recalling equation (\ref{both_PDF}) we have
\begin{eqnarray}
\bar{P}(x,t)&=&\int_0^{\infty}\frac{\eta}{D^\nu_\star\Gamma(\nu/\eta)} D^{\nu-1}
e^{-(D/D_\star)^{\eta}}\frac{1}{\sqrt{4\pi Dt}}e^{-\frac{x^2}{4Dt}}dD\nonumber\\
&=&\frac{\eta}{D^\nu_\star\Gamma(\nu/\eta)\sqrt{4\pi t}}\int_0^{\infty}D^{\nu-3/2}
e^{-D^{\eta}}e^{-\lambda D_\star/D}dD,
\label{PDF_ST}
\end{eqnarray}
where we set $\lambda=x^2/4D_\star t$. Changing the variable of integration to
$y=(D/D_\star)^\eta$ we get
\begin{eqnarray}
\bar{P}(x,t)&=&\frac{\eta}{\Gamma(\nu/\eta)\sqrt{4\pi D_\star t}}\int_0^{\infty}
y^{\frac{\nu-3/2}{\eta}}e^{-y}e^{-\lambda y^{-1/\eta}}\frac{1}{\eta}y^{\frac{1}{
\eta}-1}dy\nonumber\\
&=&\frac{1}{\Gamma(\nu/\eta)\sqrt{4\pi D_\star t}}\int_0^{\infty}y^{-1-(1/2-\nu)/
\eta}e^{-y-\lambda y^{-1/\eta}}dy.
\end{eqnarray}
With the identification
\begin{equation}
e^{-z}= H^{1,0}_{0,1}\left[z\left|\begin{array}{l}\rule[0.14cm]{1.0cm}{0.02cm}\\
(0,1)\end{array}\right.\right]
\end{equation}
with the Fox $H$-function and exploiting some (very convenient) properties of
the $H$-function \cite{prudnikov3} we then obtain
\begin{equation}
\bar{P}(x,t)=\frac{1}{\Gamma(\nu/\eta)\sqrt{4\pi D_\star t}}H^{2,0}_{0,2}\left[
\lambda\left|\begin{array}{l}\rule[0.14cm]{1.0cm}{0.02cm}\\((\nu-1/2)/\eta,1/\eta)
(0,1)\end{array}\right.\right].
\end{equation}
The Fox function is defined as a generalised Mellin-Barnes integral and has
very convenient properties under integral transformations. The Fox function
comprises a large range of special functions, including Mejer's $G$-function,
hypergeometric functions, or Bessel functions \cite{mathai-sax}. In the notation
used here the vertical line separates the argument from the function's parameters,
and the horizontal line denotes the lack of upper parameters \cite{mathai-sax}.

Recalling that $\lambda=x^2/4 D_\star t$, we finally obtain
\begin{eqnarray}
\bar{P}(x,t)&=&\frac{1}{\Gamma(\nu/\eta)\sqrt{4\pi D_\star t}}H^{2,0}_{0,2}\left[
\frac{x^2}{4D_\star t}\left|\begin{array}{l}\rule[0.14cm]{1.0cm}{0.02cm}\\((\nu-
1/2)/\eta,1/\eta)(0,1)\end{array}\right.\right]\nonumber\\
&=&\frac{1}{\Gamma(\nu/\eta)\sqrt{4\pi D_\star t}}\left(\frac{x^2}{4D_\star t}
\right)^{\nu-1/2}\nonumber\\
&&\times H^{2,0}_{0,2}\left[\frac{x^2}{4D_\star t}\left|\begin{array}{l}
\rule[0.14cm]{1.0cm}{0.02cm}\\(0,1/\eta)(-\nu+1/2,1)\end{array}\right.\right].
\label{PDF_ST_Hfunc}
\end{eqnarray}
The series expansion of this function reads \cite{mathai-sax}
\begin{eqnarray}
\nonumber
H^{2,0}_{0,2}\left[z\left|\begin{array}{l}\rule[0.14cm]{1.0cm}{0.02cm}\\(0,1/\eta)
(-\nu+1/2,1)\end{array}\right.\right]&=&\sum_{n=0}^{\infty}\frac{(-1)^n}{n!}
\Gamma(1/2-\nu-\eta n)\eta z^{\eta n}\\
&&\hspace*{-1.8cm}+\sum_{n=0}^{\infty}\frac{(-1)^n}{n!}\Gamma\left(
\frac{\nu-1/2-n}{\eta}\right)z^{1/2-\nu+n}.
\end{eqnarray}
The asymptotic behaviour is then obtained in the form \cite{mathai-sax}
\begin{equation}
\fl f_\mathrm{DD}^\mathrm{ST}(x,t)\sim\frac{1}{\Gamma(\nu/\eta)\sqrt{4\pi D_\star
t}}\left(\frac{x^2}{4D_\star t}\right)^{\frac{2\nu-\eta-1}{2(\eta+1)}}\exp\left(
-\frac{\eta+1}{\eta}\eta^{\frac{1}{\eta+1}}\left(\frac{x^2}{4D_\star t}\right)^{
\eta/(\eta+1)}\right),
\end{equation}
for $|x|\to\infty$.

\subsection{Computation via Mellin transform}
\label{app_2}

It is possible to rearrange the integral in equation (\ref{both_PDF}) as a
convolution integral,
\begin{eqnarray}
\bar{P}(x,t)&=&\int_0^{\infty}p_D(D)\frac{\exp(-\frac{(x/\sqrt{t})^2}{4D})}{
\sqrt{4\pi t}\sqrt{D}}dD\nonumber\\
&=&\frac{2}{\sqrt{4t}}\int_0^{\infty}\sqrt{D}p_D((\sqrt{D})^2)\frac{\exp(-
\frac{(x/\sqrt{D})^2}{4})}{\sqrt{\pi}}\frac{d\sqrt{D}}{\sqrt{D}}\nonumber\\
&=&\frac{1}{\sqrt{4t}}\int_0^{\infty}2\xi p_D(\xi^2)M_{1/2}\left(\frac{
\bar{x}}{\xi}\right)\frac{d\xi}{\xi},
\label{integral_4Mellin}
\end{eqnarray}
where we defined $\bar{x}=x/t^{1/2}$ and $\xi=D^{1/2}$, and $M_{1/2}$ denotes
the $M$-Wright function with parameter $\beta=1/2$ \cite{lumapa:ggBM}.
Considering the convolution formula for the Mellin transform
\begin{equation}
\int_0^{\infty}f(\xi)g\left(\frac{x}{\xi}\right)\frac{d \xi}{\xi}\quad
\overset{\mathcal{M}}{\longleftrightarrow}\quad f^M(s)g^M(s),
\label{Mellin_conv}
\end{equation}
and remembering the property
\begin{equation}
x^{\beta}f(ax^h)\quad\overset{\mathcal{M}}{\longleftrightarrow}\quad h^{-1}
a^{-(s+\beta)/h}f^M\left(\frac{s+\beta}{h}\right),
\label{mellin_prop}
\end{equation}
we compute the Mellin transform of the obtained integral in equation
(\ref{integral_4Mellin}), recovering
\begin{eqnarray}
\sqrt{4t}\bar{P}(\bar{x})&=\int_0^{\infty}2\xi p_D(\xi^2)M_{1/2}\left(\frac{
\bar{x}}{\xi}\right)\frac{d\xi}{\xi}\nonumber\\
&\overset{\mathcal{M}}{\longleftrightarrow}p^M_D\left(\frac{s+1}{2}\right)M^M_{
1/2}(s).
\label{mellin_1}
\end{eqnarray}
The Mellin transforms for the $M$-Wright function \cite{lumapa:ggBM} and the
generalised Gamma distribution \cite{mathai-sax} are known and given by
\begin{eqnarray}
M_{\beta}(x)&\overset{\mathcal{M}}{\longleftrightarrow}&\frac{\Gamma(s)}{\Gamma(
1+(s-1)/\beta)},\nonumber\\
\gamma_{(\nu,\eta)}^{gen}(x)&\overset{\mathcal{M}}{\longleftrightarrow}&D_\star^{
s-1}\frac{\Gamma((\nu+s-1)/\eta)}{\Gamma(\nu/\eta)}.
\label{mellin_transforms}
\end{eqnarray}
We can thus rewrite equation (\ref{mellin_1}) in the form
\begin{eqnarray}
\fl p^M_D \left(\frac{s+1}{2}\right)M^M_{1/2}(s)&=&D_\star^{\frac{s-1}{2}}\frac{
\Gamma\left(\frac{\nu+(s+1)/2-1}{\eta}\right)}{\Gamma\left(\frac{\nu}{\eta}\right)}
\frac{\Gamma(s)}{\Gamma\left(1+\frac{1}{2}(s-1)\right)}\\
&=&\frac{2D_\star^{\frac{s-1}{2}}}{\Gamma\left(\frac{\nu}{\eta}\right)}\frac{\Gamma
\left(\frac{\nu-1/2+s/2}{\eta}\right)\Gamma(s-1)}{\Gamma\left(\frac{s-1}{2}\right)}
\nonumber\\
&=&\frac{2D_\star^{\frac{s-1}{2}}}{\Gamma\left(\frac{\nu}{\eta}\right)}\frac{\Gamma
\left(\frac{\nu-1/2+s/2}{\eta}\right)\Gamma\left((s-1)/2+1/2\right)}{2^{1-(s-1)}
\sqrt{\pi}}\nonumber\\
&=&\frac{1}{2\sqrt{D_\star\pi}\Gamma\left(\frac{\nu}{\eta}\right)}\left(\frac{1}{4
D_\star}\right)^{-s/2}\Gamma\left(\frac{\nu-1/2}{\eta}+\frac{s/2}{\eta}\right)
\Gamma \left(\frac{s}{2}\right) .
\end{eqnarray}
Now we notice that the Mellin transform of the $H$-function is \cite{mathai-sax}
\begin{equation}
\fl H_{p,q}^{m,n}\left[ax\left|\begin{array}{l}(a_p, A_p)\\(b_q, B_q)\end{array}
\right.\right]\overset{\mathcal{M}}{\longleftrightarrow}a^{-s}\frac{\left(\prod_{
j=1}^m\Gamma(b_j+B_j s)\right)\left(\prod_{j=1}^n\Gamma(1-a_j-A_j s)\right)}{\left(
\prod_{j=m+1}^q\Gamma(1-b_j-B_j s)\right)\left(\prod_{j=n+1}^p\Gamma(a_j+A_js)
\right)}.
\label{mellin_H_func}
\end{equation}
Thus, recalling also the property of the Mellin transform in equation
(\ref{mellin_prop}) we obtain that
\begin{equation}
\sqrt{4t}\bar{P}(\bar{x})=\frac{1}{\Gamma\left(\nu/\eta\right)\sqrt{\pi D_\star}}
H_{0,2}^{2,0}\left[\frac{\bar{x}^2}{4D_\star}\left|\begin{array}{l}\rule[0.14cm]{
1.0cm}{0.02cm}\\(\frac{\nu-1/2}{\eta},\frac{1}{\eta})(0,1)\end{array}\right.\right],
\end{equation}
and finally
\begin{equation}
\bar{P}(x,t)=\frac{1}{\Gamma\left(\nu/\eta\right)\sqrt{4\pi D_\star t}}H_{0,2}^{
2,0}\left[\frac{x^2}{4D_\star t}\left|\begin{array}{l}\rule[0.14cm]{1.0cm}{0.02cm}
\\(\frac{\nu -1/2}{\eta},\frac{1}{\eta})(0,1)\end{array}\right.\right].
\label{ggBM_PDF_Mell}
\end{equation}
The result here recovered is consistent with equation (\ref{PDF_ST_Hfunc}).

\subsection{Asymptotic trend via Laplace method}
\label{app_3}

Starting again from equation (\ref{both_PDF}) it is also possible to calculate
directly the asymptotic behaviour through the Laplace method. We introduce the
new variable $y=D_\star/D$ in equation (\ref{both_PDF}),
\begin{eqnarray}
\bar{P}(x,t)&=&\frac{\eta}{D^\nu_\star\Gamma(\nu/\eta)\sqrt{4\pi t}}\int_0^{\infty}
D^{\nu-3/2}e^{-(D/D_\star)^{\eta}}e^{-\lambda D_\star/D}dD\nonumber\\ 
&=&\frac{\eta}{\Gamma(\nu/\eta)\sqrt{4\pi D_\star t}}\int_0^{\infty}y^{-\nu-1/2}
e^{-y^{-\eta}-\lambda y}dy
\end{eqnarray}
Now the integral looks like a Laplace integral of the form
\begin{equation}
I(\lambda)=\int_0^{\infty}f(y) e^{-\lambda y}dy.
\label{Laplace_int}
\end{equation}
In order to apply the Laplace method we need $f(0)\ne0$ which is not our case since
$f(0)=0$ together with all its derivatives. Thus, to evaluate the asymptotics, we
define the maximum of the function,
\begin{equation}
\phi(y)=-\lambda y-y^{-\eta},
\end{equation}
which is located at $y_m=(\eta/\lambda)^{1/\eta+1}$. Introducing of the new
variable $\bar{z}=z\eta^{\frac{1}{\eta+1}}$ the integral becomes
\begin{eqnarray}
I(\lambda)&=&\lambda^{\frac{2\nu-1}{2(\eta+1)}}\int_0^{\infty}\bar{z}^{\frac{2
\nu-1}{2(\eta+1)}}\exp\left[-\lambda^{\frac{\eta}{\eta+1}}(\bar{z}^{-\eta}+\bar{
z})\right]d\bar{z}\nonumber\\ 
&=&\lambda^{\frac{2\nu-1}{2(\eta+1)}}\int_0^{\infty}\bar{z}^{\frac{2\nu-1}{2(\eta
+1)}}e^{\bar{\lambda}f(\bar{z})}d\bar{z},
\label{I_lambda}
\end{eqnarray}
where we defined $f(\bar{z})=-\bar{z}-\bar{z}^{-\eta}$ and $\bar{\lambda}=\lambda^{
\eta/(\eta+1)}$. Now the standard Laplace method can be applied considering that
the function $f(\bar{z})$ reaches its maximum at $\bar{z}_m=\eta^{1/(\eta+1)}$,
such that
\begin{eqnarray}
I(\lambda)&=&\lambda^{\frac{2\nu-1}{2(\eta+1)}}\bar{z}_m^{\frac{2\nu+1}{2(\eta+1)}}
e^{\bar{\lambda}f(\bar{z}_m)}\sqrt{\frac{2\pi}{\bar{\lambda}|f''(\bar{z}_m)|}}
\nonumber\\
&=&\sqrt{\frac{2\pi}{\eta+1}}\eta^{\frac{2\nu+\eta+1}{2(\eta+1)^2}}\lambda^{\frac{
2\nu-\eta-1}{2(\eta+1)}}\exp\left[-\frac{\eta+1}{\eta}\eta^{\frac{1}{\eta+1}}
\lambda^{\frac{\eta}{\eta+1}}\right].
\end{eqnarray}
This finally leads to
\begin{equation}
\fl \bar{P}(x,t)\simeq\frac{\eta^{\frac{2\nu+\eta+1}{2(\eta+1)^2}+1}}{\Gamma(\nu/
\eta)\sqrt{4\pi D_\star t}}\sqrt{\frac{2\pi}{\eta +1}}\left(\frac{x^2}{4D_\star t}
\right)^{\frac{2\nu-\eta-1}{2(\eta+1)}}\exp\left[-\frac{\eta+1}{\eta}\eta^{\frac{
1}{\eta+1}}\left(\frac{x^2}{4D_\star t}\right)^{\frac{\eta}{\eta+1}}\right],
\end{equation}
for $|x|\to\infty$. This result is, up to a numerical prefactor, identical to the
asymptotic behaviour obtained in (\ref{app_1_Hfunc}).

\section*{References}

\bibliographystyle{iopart-num}

\end{document}